\documentclass[paper,notoc]{JHEP3}

\usepackage{epsfig}
\usepackage{graphicx}
\usepackage{feynmp}

\usepackage{amsmath}
\usepackage{amsfonts}
\usepackage{amssymb}
\usepackage{graphicx}

\newcommand{\ord}{\begin{cal}O\end{cal}}

\newcommand\sss{\scriptscriptstyle}
\newcommand\as{\alpha_{\sss S}}
\newcommand\tp{\tilde p}

\def\de{\delta}
\def\beq{\begin{equation}}
\def\eeq{\end{equation}}
\def\beqa{\begin{eqnarray}}
\def\eeqa{\end{eqnarray}}

\def\bsp#1\esp{\begin{split}#1\end{split}}
\def\eqn#1{eq.~(\ref{#1})}

\newenvironment{sloppyequation}[0]{\sloppy\begin{flushleft}\hspace*{0.75cm}\(}{\)\end{flushleft}\fussy}
\newcommand{\beqsloppy}{\begin{sloppyequation}}
\newcommand{\eeqsloppy}{\end{sloppyequation}}

\font\cyr=wncyr8
\newcommand{\sha}{{\mbox{\cyr X}}}
\newfont{\scyr}{wncyr10 scaled 550}
\newcommand{\ssha}{\mbox{\bf \scyr X}}

\newcommand{\cL}{\begin{cal}L\end{cal}}

\newcommand{\cN}{\begin{cal}N\end{cal}}

\newcommand{\cP}{\begin{cal}P\end{cal}}

\newcommand{\cZ}{\begin{cal}Z\end{cal}}

\title{The BFKL equation, Mueller-Navelet jets and single-valued 
harmonic polylogarithms}

\author{Vittorio Del Duca\\
INFN Laboratori Nazionali di Frascati\\
00044 Frascati (Roma), Italy\\
       E-mail: \email{delduca@lnf.infn.it}}

\author{Lance J. Dixon\\
SLAC National Accelerator Laboratory\\
Stanford University, Stanford CA 94309, USA\\
E-mail: \email{lance@slac.stanford.edu}}

\author{Claude Duhr\\
Institut f\"ur Theoretische Physik, ETH Z\"urich\\
Wolfgang-Paulistrasse 27, CH 8093, Z\"urich, Switzerland\\
  Institute for Particle Physics Phenomenology, University of Durham\\
  Durham, DH1 3LE, U.K.\\
E-mail: \email{duhrc@itp.phys.ethz.ch}}

\author{Jeffrey Pennington\\
SLAC National Accelerator Laboratory\\ 
Stanford University, Stanford CA 94309, USA\\
E-mail: \email{jpennin@stanford.edu}}

\abstract{We introduce a generating function for the coefficients of the
leading logarithmic BFKL Green's function in transverse-momentum space,
order by order in $\as$, in terms of single-valued harmonic polylogarithms. 
As an application, we exhibit fully analytic azimuthal-angle
and transverse-momentum distributions for Mueller-Navelet jet cross
sections at each order in $\as$.  We also provide a generating function for the total cross section valid to any number of loops.}

\preprint{SLAC--PUB--15706\\IPPP/13/76\\DCPT/13/152}

\begin{document}

\catcode`\@=11
\font\manfnt=manfnt
\def\Watchout{\@ifnextchar [{\W@tchout}{\W@tchout[1]}}
\def\W@tchout[#1]{{\manfnt\@tempcnta#1\relax%
  \@whilenum\@tempcnta>\z@\do{%
    \char"7F\hskip 0.3em\advance\@tempcnta\m@ne}}}
\let\foo\W@tchout
\def\dubious{\@ifnextchar[{\@dubious}{\@dubious[1]}}
\let\enddubious\endlist
\def\@dubious[#1]{%
  \setbox\@tempboxa\hbox{\@W@tchout#1}
  \@tempdima\wd\@tempboxa
  \list{}{\leftmargin\@tempdima}\item[\hbox to 0pt{\hss\@W@tchout#1}]}
\def\@W@tchout#1{\W@tchout[#1]}
\catcode`\@=12

\section{Introduction}
\label{sec:intro}

In the limit in which the squared center-of-mass energy is much greater than
the momentum transfer, $s \gg |t|$, any QCD scattering process is dominated
by gluon exchange in the $t$ channel. Building upon this fact,
the Balitsky-Fadin-Kuraev-Lipatov (BFKL) theory models
strong-interaction processes with two large and disparate scales,
by resumming the radiative corrections to parton-parton
scattering. This is achieved to leading logarithmic accuracy, in
$\ln(s/|t|)$, through the BFKL 
equation~\cite{Kuraev:1976ge,Kuraev:1977fs,Balitsky:1978ic}, 
{\it i.e.}~an integral equation which describes the evolution of the
$t$-channel gluon propagator in transverse-momentum space and Mellin
moment space. The integral equation is obtained by computing the one-loop
leading-logarithmic corrections to the gluon exchange in the $t$ channel.
They are formed by a real correction -- the emission of a gluon along the 
ladder~\cite{Lipatov:1976zz} -- and a virtual correction --
the so-called one-loop Regge trajectory.
The BFKL equation is then obtained by iterating these
one-loop corrections to all orders in $\as$, to leading logarithmic 
accuracy. For $\as \ln(s/|t|) \gg 1$, one can provide an analytic solution
to the BFKL equation in the saddle-point approximation~\cite{Kuraev:1977fs}.
The next-to-leading logarithmic (NLL) corrections to the BFKL equation
are also known~\cite{Fadin:1998py,Camici:1997ij,Ciafaloni:1998gs}.

The amplitude for any QCD scattering process factorizes at
leading-logarithmic accuracy into a gauge-invariant effective amplitude
formed by two scattering centers, the leading order impact factors,
connected by a Reggeized gluon exchanged in the crossed channel. The leading 
order impact factors depend on the particular scattering process.
The Reggeized gluon exchange in the $t$ channel is process-independent,
and it is described by the BFKL equation.
Since the exchange of a gluon in the $t$ channel is required,
for an arbitrary scattering the leading-order term of the BFKL resummation
is often contained in the higher-order terms of the expansion in $\as$.
(For example, in Drell-Yan production gluon exchange in the
$t$ channel occurs at $\ord(\as^2)$, two orders above the leading order
term of the expansion in $\as$.)  For dijet production in hadron
collisions, the leading order term in BFKL resummation occurs
already in the leading order term of the expansion in $\as$. 
In this respect, dijet production in hadron collisions is the simplest
process in which to consider BFKL resummation.

Long ago, Mueller and Navelet~\cite{Mueller:1986ey} suggested to look for
evidence of BFKL evolution by measuring the dijet cross section at
hadron colliders as a function of the hadronic
centre-of-mass energy $\sqrt{S}$, at fixed momentum fractions 
$x_{a,b}$ of the incoming partons, and fixed minimum jet
transverse momentum.  This is equivalent to measuring the
rates as a function of the rapidity interval $\Delta y$ between 
the jets. In fact, at large enough rapidities, 
the rapidity interval is well approximated by 
$\Delta y\simeq \ln(s/|t|)$, where $s=x_a x_b S$ and 
$|t|\simeq {p_1}_{_\perp} {p_2}_{_\perp}$, 
with ${p_1}_{_\perp}$ and ${p_2}_{_\perp}$ the transverse momenta of the two jets. 
Thus, since the cross section tends to peak at the smallest available 
transverse momenta, $\Delta y$ grows as $\ln S$ at fixed $x_{a,b}$.
A measurement of the dijet rate at different centre-of-mass energies
allows in principle for a measurement of the leading eigenvalue appearing
in BFKL resummation, {\it i.e.}~of the BFKL intercept.
Conversely, in a fixed-energy collider, $\Delta y$ grows with $x_{a,b}$
at fixed $S$. Then evidence of the BFKL resummation may be looked for
by studying dijet production cross sections at large rapidity 
intervals~\cite{Mueller:1986ey,DelDuca:1993mn,Stirling:1994he}, 
as well as the distribution in azimuthal angle between the two 
jets with the largest rapidity separation~\cite{DelDuca:1993mn,%
Stirling:1994he,DelDuca:1994ng,DelDuca:1994fx,Orr:1997im}.

The initial computations of radiative corrections to these observables
were performed at leading-logarithmic (LL) accuracy in $\Delta y$.
Improvements in several directions then followed, most notably in
treating the BFKL resummation numerically using Monte Carlo event 
generators~\cite{Orr:1997im,Schmidt:1996fg,Andersen:2001kta,Andersen:2006sp},
and in including the NLL BFKL corrections to dijet 
production~\cite{Colferai:2010wu}.  A phenomenological analysis of
dijet production at large $\Delta y$ at NLL accuracy can be found in
ref.~\cite{Ducloue:2013hia}.
The dijet cross section as a function of the rapidity interval $\Delta y$
has been measured experimentally by the D0 experiment~\cite{Abbott:1999ai}
at the Tevatron at different centre-of-mass energies, and 
by the ATLAS~\cite{Aad:2011jz}
and CMS~\cite{Chatrchyan:2012pb} experiments at the LHC
at a fixed centre-of-mass energy $\sqrt{S} = 7$~TeV.
The azimuthal-angle distributions of pairs of jets with wide rapidity
separations have been measured by the D0
experiment~\cite{Abachi:1996et}, and quite recently by the
CMS experiment~\cite{CMSazimuthal}.

In this paper, we use Mueller-Navelet jets as a template to examine
the analytic dependence of the LL
BFKL equation on the transverse momenta
of the partons that delimit the BFKL ladder.  The partons' transverse
momenta equal the transverse momenta of the two tagging jets in dijet
production at large $\Delta y$.  By writing the dijet cross section as an
expansion in $\as \Delta y$, Mueller and Navelet were able to integrate
analytically
over the transverse momenta of the two tagging jets in the first few
orders of the expansion. However, for the fully differential dijet
cross section only a numerical solution could be provided, through either
a direct integration or a Monte Carlo event generator.  Recently it
was found~\cite{Dixon:2012yy,Pennington:2012zj} that the six-gluon
amplitude in $\mathcal{N}=4$ super-Yang-Mills theory in multi-Regge kinematics
can be described purely in terms of a class of mathematical functions
known as single-valued harmonic (or multiple)
polylogarithms (SVHPLs)~\cite{BrownSVHPLs}.  Motivated by these developments, 
in this paper we use these functions to write the BFKL ladder in
transverse momentum space, and thus the fully differential dijet cross
section, as explicit analytic functions of the
transverse momenta of the two tagging jets.  More precisely, we
introduce a generating function for
the coefficients of the LL BFKL Green's function,
which allows us to express the Green's function, order by order in
perturbation theory, as a combination of polylogarithmic functions
of one complex variable --- the complexified transverse momentum ---
that are single-valued in the complex plane.

The paper is organized as follows:  In Section~\ref{sec:mn}, we recall
the BFKL Green's function and the Mueller-Navelet dijet cross section.
In Section~\ref{sec:SVHPLs} we introduce our toolbox, the
single-valued harmonic polylogarithms. In Section~\ref{sec:BFKL_all_orders},
we introduce a generating function for the coefficients of the 
BFKL Green's function in transverse-momentum space at leading-logarithmic
accuracy, and we display the coefficients explicitly up to the seventh loop.
In Section~\ref{sec:mnjets}, we sketch how our results for the BFKL
Green's function can easily be integrated over the azimuthal angle
or the transverse momentum, and we provide fully analytic azimuthal-angle
and transverse-momentum distributions for the Mueller-Navelet jet cross section
in terms of harmonic polylogarithms up to the sixth loop, as well as a generating function for the transverse momentum distribution to any number of loops. Finally, in the same section we explicitly perform the integration of the transverse momentum distribution to obtain the total cross section, and we present a generating function for the Mueller-Navelet coefficients to any number of loops, as well as explicit results in terms of multiple zeta values up to 13 loops.
In Section~\ref{sec:concl}, we draw our conclusions.


\section{The BFKL equation and Mueller-Navelet jets}
\label{sec:mn}

The cross section for dijet production in the high-energy limit
is dominated by gluon exchange in the crossed channel.
Thus, at leading order in $\as$ the functional form of the QCD amplitudes
for gluon-gluon, gluon-quark or quark-quark scattering is the
same; they differ only by the color strength in the parton-production
vertices.  Hence the cross section takes the following factorized
form,
\begin{equation}
{d\sigma\over d p_{1_\perp}^2 d p_{2_\perp}^2 d\phi_{jj} 
dy_1 dy_2}\, =\,
x_a^0 f_{\rm eff}(x_a^0,\mu_F^2)\, x_b^0 f_{\rm eff}(x_b^0,\mu_F^2)\,
{d\hat\sigma_{gg}\over d p_{1_\perp}^2 d p_{2_\perp}^2
d\phi_{jj} }\, ,\label{mrfac}
\end{equation}
where $1$ and $2$ label the forward and backward outgoing jets, respectively, 
$p_{i\bot}$ and $y_i$ are the jet transverse momenta and rapidities, and
$\phi_{jj}$ denotes the azimuthal angle between the jets.
In the high-energy limit, the parton momentum fractions are given by,
\beq
x_a^0 = \sqrt{p_{1_\perp}^2\over S}\, e^{y_1}\,,\qquad \qquad
x_b^0 = \sqrt{p_{2_\perp}^2\over S}\, e^{-y_2}\,,\label{nkin0}
\eeq
and the effective parton distribution functions are
\begin{equation}
f_{\rm eff}(x,\mu_F^2) = G(x,\mu_F^2) + {4\over 9}\sum_f
\left[Q_f(x,\mu_F^2) + \bar Q_f(x,\mu_F^2)\right], \label{effec}
\end{equation}
where the sum is over the quark flavors, and $\mu_F$ is the factorization
scale. 

The gluon-gluon scattering cross section in the high-energy limit becomes,
\beq
\frac{d\hat\sigma_{gg}}{dp_{1\bot}^2dp_{2\bot}^2d\phi_{jj}}
= \frac{\pi}{2}\,\left[\frac{C_A\as}{p_{1\bot}^2}\right]
\,f(\vec{q}_{1\bot},\vec{q}_{2\bot},\Delta y)
\,\left[\frac{C_A\as}{p_{2\bot}^2}\right]\,,
\label{cross}
\eeq
where $\Delta y=y_1-y_2$ is the rapidity difference between the two jets. 
The variables $\vec{q}_{i\bot}$ that enter the Green's function
$f(\vec{q}_{1\bot},\vec{q}_{2\bot},\Delta y)$ are related to the
transverse momenta of the jets by $\vec{q}_{1\bot} =-\vec{p}_{1\bot}$
and $\vec{q}_{2\bot} =\vec{p}_{2\bot}$. Fixing
\beq
\eta\equiv \displaystyle{\frac{C_A\,\as}{\pi} }\,\Delta y\,,
\eeq
at leading logarithmic accuracy the Green's function is given by
\beq
f(\vec{q}_{1\bot},\vec{q}_{2\bot},\Delta y)
= \frac{1}{(2\pi)^2\, \sqrt{q_{1\bot}^2\,q_{2\bot}^2} }
\,\sum_{n=-\infty}^{+\infty} e^{in\phi}
\,\int_{-\infty}^{+\infty}d\nu\,\left(\frac{q_{1\bot}^2}{q_{2\bot}^2}\right)^{i\nu}
\,e^{\eta\, \chi_{\nu,n}}\,,
\label{green}
\eeq
where $q_{i\bot}^2 = |\vec{q}_{i\bot}|^2$,
and $\phi$ is the angle between $\vec{q}_{1\bot}$ and $\vec{q}_{2\bot}$
(and therefore $\phi = \pi-\phi_{jj}$).  The exponent in \eqn{green}
is given by $\eta\, \chi_{\nu,n} = \Delta y\,\omega(\nu,n)$, where
\beq
\omega(\nu,n) = \frac{C_A\as}{\pi}\,\chi_{\nu,n}\,,
\eeq
is the LL BFKL eigenvalue, with
\beq
 \chi_{\nu,n} = -2\gamma_E
-\psi\left(\frac{1}{2}+\frac{|n|}{2}+i\nu\right)
-\psi\left(\frac{1}{2}+\frac{|n|}{2}-i\nu\right)\,.
\eeq

\subsection{The saddle-point approximation}
\label{sec:saddlept}

For $\eta \gg 1$, we may perform a saddle-point approximation to
the integral over $\nu$ in the gluon Green's function~(\ref{green}).
The saddle point is near $\nu=0$.  We use the small-$\nu$ expansion of the
BFKL eigenvalue, \eqn{omegacoeff},
\beq
\chi_{\nu,n} = 2 \left(a_{0n} + a_{1n} \nu^2 + \cdots \right)\,,
\eeq
with the coefficients given in \eqn{fine}. Then we can write the gluon
Green's function~(\ref{green}) as,
\beq
f(\vec{q}_{1\bot},\vec{q}_{2\bot},\Delta y) 
= \frac1{4\pi\, \sqrt{q_{1\bot}^2\,q_{2\bot}^2} }
\,\sum_{n=-\infty}^{+\infty}e^{in\phi}\,
\frac{\exp\left(2a_{0n} \eta - \displaystyle
\frac{\ln^2\frac{q_{1\bot}^2}{q_{2\bot}^2}}{4(-2a_{1n})\eta} \right) }
{\sqrt{\pi (-2a_{1n})\eta}}\, .
\label{saddle}
\eeq
Note that $a_{1n}$, as given in \eqn{fine}, is always negative;
thus the saddle-point approximation is well defined.
Integrating out the azimuthal angle, which singles out the $n=0$
contribution, and using $a_{00}$ and $a_{10}$ as given in \eqn{chi0}, 
we obtain the usual saddle-point approximation of the BFKL Green's function.

\subsection{The azimuthal angle distribution}
\label{sec:azim}

Integrating out the jet transverse momenta over
$E_{_\perp} \le p_{_\perp} < \infty$, \eqn{cross} becomes
\begin{equation}
{d\hat\sigma_{gg}\over d\phi_{jj} }\ = \frac{\pi (C_A\as)^2}{2 E_\bot^2} 
\left[\delta(\phi_{jj}-\pi)
+ \sum_{k=1}^\infty  
\left(\sum_{n=-\infty}^{\infty} \frac{e^{in\phi}}{2\pi} f_{n,k}\right)
\eta^k \right]\, ,\label{intxsect}
\end{equation}
with
\beq
f_{n,k} = {1\over 2\pi}\, \frac1{k!}\, 
\int_{-\infty}^{\infty} d\nu\, \frac{\chi_{\nu,n}^k}{\nu^2 + \frac{1}{4}}\,.
\label{fnk}
\eeq
The Born term has been singled out by performing the integral in
$f_{n,0}$, which yields $f_{n,0} = 1$, and furthermore by using,
\beq
{1\over 2\pi} \sum_{n=-\infty}^{\infty} e^{in\phi} = \delta(\phi_{jj}-\pi)\,.
\eeq
In the limit $\as \Delta y\to 0$, \eqn{intxsect} reduces to the
azimuthal distribution at leading order.  Note that by using the
recursive formula (\ref{recurs2}) for the $\chi_{\nu,n}$ coefficients,
we can obtain a recursive formula for the Fourier coefficients $f_{n,k}$,
in terms of a one-dimensional integral over $\nu$.

\subsection{The Mueller-Navelet dijet cross section}
\label{sec:azimn}

When the azimuthal angle is integrated out over the full range
$0\le\phi_{jj}\le 2\pi$, only the zero mode of \eqn{intxsect} survives,
\beq
\int_0^{2\pi} d\phi_{jj} e^{in\phi} = 2\pi \delta_{n,0}\, ,
\eeq
and we obtain the Mueller-Navelet dijet cross section,
\beq
\hat\sigma_{gg} = \frac{\pi (C_A\as)^2}{2 E_\bot^2}
\, \sum_{k=0}^\infty f_{0,k}\, \eta^k\,.
\label{eq:mnexp}
\eeq
Mueller and Navelet~\cite{Mueller:1986ey} computed analytically
the first few coefficients of the expansion (\ref{eq:mnexp}),
\beq\bsp
\label{MNtotalcross}
f_{0,0} &\,= 1\,, \\
f_{0,1} &\,= 0\,,  \\
f_{0,2} &\,= 2\zeta_2\,, \\
f_{0,3} &\,= -3\zeta_3\,, \\
f_{0,4} &\,= \frac{53}6\, \zeta_4\,, \\
f_{0,5} &\,= -\frac1{12}\, (115\zeta_5 + 48\zeta_2\zeta_3)\,.
\esp\eeq

However, for the fully differential dijet cross section~(\ref{cross}),
so far only a numerical solution could be provided, through either a
direct integration or a Monte Carlo event generator.


\section{Single-valued harmonic polylogarithms}
\label{sec:SVHPLs}

In this section we give a short review of the main tools that allow us
to solve the Green's function perturbatively, namely harmonic
polylogarithms (HPLs) and their single-valued analogues. We start by
reviewing the multi-valued case, and we then briefly recall the
construction of the single-valued analogues of
ref.~\cite{BrownSVHPLs} (see also ref.~\cite{Brown:2013gia}).

Harmonic polylogarithms were first introduced in the physics
literature in ref.~\cite{Remiddi:1999ew}. They are defined iteratively
through the differential equations\footnote{We only consider here the
  case of poles at 0 or 1.}
\begin{equation}\label{eq:HPLdef_1}
\frac{d}{dz}H_{0\,\omega}(z) 
= \frac{H_{\omega}(z)}{z}
\quad\quad\text{and}\quad\quad
\frac{d}{dz}H_{1\,\omega}(z) = \frac{H_{\omega}(z)}{1-z}\,,
\end{equation}
where $\omega$ denotes any word formed out of the letters `0' and
`1'. The length $|\omega|$ of the word $\omega$ is called the
\emph{weight} of $H_\omega(z)$. The solutions of the differential
equation are subject to the constraints,
\begin{equation}
H(z)=1,\quad\quad 
H_{\vec{0}_n}(z)=\frac{1}{n!}\ln^nz,
\quad\quad\text{and}\quad\quad 
\lim_{z\rightarrow 0}H_{\omega\neq \vec{0}_n}(z)=0\,.
\end{equation}
It is easy to check that the solutions to these differential equations
are given by the iterated integrals
\beq
H_{a\omega}(z) = \int_0^z dt\,f_a(t)\,H_\omega(t)\,,
\label{eq:Hintdef}
\eeq
with 
\beq
f_0(z) = \frac{1}{z}\quad\quad\text{and}\quad\quad 
f_1(z) = \frac{1}{1-z}\,.
\eeq

HPLs enjoy many properties. In particular, they form a shuffle algebra,
{\it i.e.},
\begin{equation}
\label{eq:HPL_shuffle}
H_{\omega_1}(z)\,H_{\omega_2}(z) = \sum_{{\omega}\in{\omega_1}\ssha {\omega_2}}H_{\omega}(z)\,,
\end{equation}
where ${\omega_1}\sha{\omega_2}$ is the set of mergers, or shuffles,
of the sequences $\omega_1$ and $\omega_2$; each element of the set is
an interleaving of the two sequences such that the orderings of the letters
inside $\omega_1$, and of those inside $\omega_2$, are preserved.
Furthermore, the HPLs contain the classical polylogarithms as special cases,
\begin{equation}
\label{eq:Li_int}
H_{\scriptsize\underbrace{0\ldots 0}_{m-1}1}(z)
={\rm Li}_{m} (z) = \int_0^z dt\; \frac{{\rm Li}_{m-1}(t)}{t}\,,
\end{equation}
with ${\rm Li}_{1} (z) = -\ln(1-z)$.

Due to the singularities in the differential
equations~\eqref{eq:HPLdef_1}, HPLs in general define multi-valued
functions on the punctured complex plane $\mathbb{C}/\{0,1\}$. In
ref.~\cite{BrownSVHPLs} it was shown that for every HPL $H_\omega(z)$
there is a function $\mathcal{L}_\omega(z)$ that is real-analytic and
single-valued on $\mathbb{C}/\{0,1\}$ and that satisfies the same
properties as the ordinary HPLs.\footnote{We will drop the explicit
$\bar{z}$ argument from $\mathcal{L}_\omega(z,\bar{z})$ henceforth.}
That is, the $\cL_\omega(z)$ satisfy the differential equations
\begin{equation}\label{eq:Lzdiffeq}
\frac{\partial}{\partial z}\cL_{0\,\omega}(z) 
= \frac{\cL_{\omega}(z)}{z} 
\quad\quad\text{and}\quad\quad
\frac{\partial}{\partial z}\cL_{1\,\omega}(z) = \frac{\cL_{\omega}(z)}{1-z}\,,
\end{equation}
subject to the conditions
\begin{equation}\label{eq:cL_conditions}
\cL(z)=1\,,\quad\quad 
\cL_{\vec{0}_n}(z)=\frac{1}{n!}\ln^n|z|^2
\quad\quad\text{and}\quad\quad
\lim_{z\rightarrow 0}\cL_{\omega\neq \vec{0}_n}(z)=0\,.
\end{equation}
In addition, the SVHPLs $\mathcal{L}_\omega(z)$ also form a shuffle algebra, 
\begin{equation}
\mathcal{L}_{\omega_1}(z)\,\mathcal{L}_{\omega_2}(z)
= \sum_{{\omega}\in{\omega_1}\ssha {\omega_2}}\mathcal{L}_{\omega}(z)\,.
\label{eq:cL_shuffle}
\end{equation}
SVHPLs can be explicitly expressed as combinations of ordinary HPLs
such that all the branch cuts cancel. In order to understand how these
combinations can be constructed, and also to understand the solution
for the BFKL Green's function in Section~\ref{sec:BFKL_all_orders}, it
is useful to first get a new viewpoint on ordinary HPLs.

It is clear from the previous discussion that for every word
$\omega$ formed out of the letters `0' and `1' there is a harmonic
polylogarithm $H_\omega(z)$. Let us therefore denote the set of all
words formed out of the non-commutative variables $x_0$ and $x_1$ by
$X^*$, and let $\mathbb{C}\langle X\rangle$ be the complex vector
space generated by the elements in $X^*$, {\it i.e.}, the vector space of
all formal $\mathbb{C}$-linear combinations of elements in $X^*$.
$\mathbb{C}\langle X\rangle$ can be turned into an algebra by
equipping it with the concatenation of words as multiplication. The
set of differential equations~\eqref{eq:HPLdef_1} can then be conveniently
summarized by a single differential equation for a generating function
$L(z)$, often referred to as a Knizhnik-Zamolodchikov (KZ)
equation~\cite{Knizhnik:1984nr}
\beq
\frac{d}{dz}L(z)
= \left(\frac{x_0}{z}+\frac{x_1}{1-z}\right)\,L(z)\,,
\qquad L(z) \in \mathbb{C}\langle X\rangle\,.
\eeq
It is easy to see that a solution to the KZ equation is given by
\beq
L(z) = \sum_{\omega\in X^*}H_\omega(z)\,\omega\,,
\eeq
where obviously for the empty word $e$ we have $H_e(z)=H(z)=1$. We
can then define the single-valued analogues of the ordinary HPLs by
constructing a single-valued generating-function solution
\beq\label{eq:cL(z)}
\mathcal{L}(z) = \sum_{\omega\in X^*}\mathcal{L}_\omega(z)\,\omega
\eeq
of the analogous KZ equation, 
\beq
\frac{\partial}{\partial z}\mathcal{L}(z)
= \left(\frac{x_0}{z}+\frac{x_1}{1-z}\right)\,\mathcal{L}(z)\,,
\qquad \mathcal{L}(z) \in \mathbb{C}\langle X\rangle\,,
\eeq
subject to the conditions~\eqref{eq:cL_conditions}. We briefly review the
construction of $\cL(z)$ of ref.~\cite{BrownSVHPLs} in the rest of this
section. 

We start by defining a second alphabet $\{y_0,y_1\}$ (and a set of
words $Y^*$) and a map $^ \sim: Y^*\rightarrow Y^*$ as the operation
that reverses words. The alphabets $\{x_0, x_1\}$ and $\{y_0,y_1\}$
are not independent but they are related by the following relations,
\beq\bsp
\label{y_alph}
y_0&\,=\,x_0\\
\tilde{Z}(y_0,y_1)y_1\tilde{Z}(y_0,y_1)^{-1} &\,=\, Z(x_0,x_1)^{-1}x_1 Z(x_0,x_1)\,,
\esp\eeq
where $Z(x_0,x_1)$ denotes the Drinfel'd associator, defined by the series,
\begin{equation}
Z(x_0,x_1) = \sum_{\omega\in X^*} \zeta(\omega)\omega   = L(1),
\end{equation}
where $\zeta(\omega)=H_\omega(1)$ for $\omega\neq x_1$ and
$\zeta(x_1)=0$.  The $\zeta(\omega)$ are regularized by the shuffle
algebra~\cite{ShuffleReg}; that is, we use the shuffle algebra to define the naively divergent
cases.  Using a collapsed notation for $\omega$, in which zero entries
are removed according to $\vec{0}_{m-1}1\to m$, these
$\zeta(\omega)$ are the familiar multiple zeta values.  The inversion
operator is to be understood as a formal series expansion in the
weight $|\omega|$ (also known as the length of the word $\omega$).

We can solve eq.~\eqref{y_alph} iteratively in the
length of the word. This yields a series expansion for $y_1$.  Next,
let $\phi:Y^*\rightarrow X^*$ be the map that renames $y$ to $x$,
{\it i.e.}~$\phi(y_0)=x_0$ and $\phi(y_1)=x_1$, define the generating
functions
\beq\label{eq:LXY0}
L_X(z)\,=\,\sum_{\omega\in X^*}H_\omega(z)\omega \,, \qquad
\tilde{L}_Y(\bar{z})\,=\,\sum_{\omega\in Y^*}H_{\phi(\omega)}(\bar{z})\tilde{\omega} 
\,,
\eeq
where $\bar z$ denotes the complex conjugate of $z$. It was shown in
ref.~\cite{BrownSVHPLs} that the generating function~\eqref{eq:cL(z)}
for the single-valued analogues of HPLs is then obtained by
\beq\label{eq:cL_construction}
\cL(z) = \sum_{\omega\in X^*}\cL_\omega(z)\,\omega = L_X(z)\,\tilde{L}_Y(\bar{z})\,.
\eeq
Expanding the right-hand side of eq.~\eqref{eq:cL_construction}, the
coefficient of each word $\omega\in X^*$ is a combination of HPLs such
that the branch cuts cancel, {\it i.e.}~it is single-valued on
$\mathbb{C}/\{0,1\}$.


\section{The BFKL equation and single-valued harmonic polylogarithms}
\label{sec:BFKL_all_orders}

We now apply the ideas of the previous section to the BFKL Green's
function. In particular, we provide (at least conjecturally) a
generating function for the coefficients appearing in the perturbative
expansion of the BFKL Green's function. The Green's function is a
function of the (two-dimensional) transverse momenta $p_{1\bot}$ and
$p_{2\bot}$. For the rest of the discussion, it turns out to be
convenient to encode the information on each transverse momentum by a
single complex number, $p_{k\bot}\to {\tilde p_k} = p^x_k + i
p^y_k$. Furthermore, we introduce a complex variable $w$ by
\beq
w = \frac{\tp_1}{\tp_2} {\rm~~~~and~~~~} 
w^\ast = \frac{\tp_1^\ast}{\tp_2^\ast}\,,
\label{eq:w}
\eeq
such that
\beq
|w|^2 = \frac{|\tp_1|^2}{|\tp_2|^2} =
\frac{p_{1\bot}^2}{p_{2\bot}^2} = \frac{q_{1\bot}^2}{q_{2\bot}^2}
{\rm~~~~and~~~~}
\left(\frac{w}{w^\ast}\right)^{1/2} 
= e^{-i\phi_{jj}} = -e^{i\phi}\,.
\label{eq:variab}
\eeq

The Green's function can be expanded into a power series in $\eta$,
\beq
f(\vec{q}_{1\bot},\vec{q}_{2\bot},\Delta y) 
= \frac{1}{2}\delta^{(2)}(\vec{q}_{1\bot}-\vec{q}_{2\bot})
+\frac{1}{2\pi\, \sqrt{q_{1\bot}^2\,q_{2\bot}^2} }
\,\sum_{k=1}^\infty\eta^k \,f_k(w,w^\ast)\,,
\label{eqn:greensvhpl}
\eeq
where the coefficients $f_k$ are given by the inverse
Fourier-Mellin transform,
\beq
f_k(w,w^\ast) = \frac{1}{k!}
\sum_{n=-\infty}^{+\infty}(-1)^n\,\left(\frac{w}{w^\ast}\right)^{n/2}
\int_{-\infty}^{+\infty}\frac{d\nu}{2\pi}\,|w|^{2i\nu}\,\chi_{\nu,n}^k\,.
\label{eq:fmc}
\eeq
These coefficients should be real-analytic functions of $w$; that is,
they should have a unique, well-defined value for every ratio
of the magnitudes of the two jet transverse momenta and angle between
them. 

The invariance of \eqn{eq:fmc} under $n\leftrightarrow-n$ and
$\nu\leftrightarrow-\nu$ implies that $f_k$ is invariant under
conjugation and inversion of $w$:
\beq\label{eq:fk_symmetry}
f_k(w,w^\ast) = f_k(w^\ast,w) = f_k(1/w,1/w^\ast)\,.
\eeq
In other words, the perturbative coefficients are eigenfunctions under
the action of the $\mathbb{Z}_2\times\mathbb{Z}_2$ symmetry generated by
\beq\label{eq:Z2xZ2}
(w,w^\ast) \leftrightarrow (w^\ast,w) {\rm~~~~and~~~~}
(w,w^\ast) \leftrightarrow (1/w,1/w^\ast)\,.
\eeq
From \eqn{eq:w} we see that a special point in the $(w,w^\ast)$ plane
is at $w = w^\ast = -1$.  This is the configuration of Born kinematics,
where the two jets have equal and opposite transverse momentum.
A second preferred point is the origin, $w = w^\ast = 0$, when one jet
has much smaller transverse momentum than the other jet.
The point at infinity is related to the origin by the inversion symmetry,
while $w = w^\ast = -1$ is a fixed point of the
$\mathbb{Z}_2\times\mathbb{Z}_2$ symmetry~\eqref{eq:Z2xZ2}.
We anticipate that the SVHPLs $\cL_\omega(z,\bar{z})$ will play a role.
However, their poles are at $z=0$ and 1, not $w=0$ and $-1$, 
so we will need to identify $(z,\bar{z})=(-w,-w^\ast)$.

In the rest of this section we argue that one can use SVHPLs
to write down a
generating function for the perturbative coefficients $f_k(w,w^\ast)$
to all orders in the expansion parameter $\eta$. In order to present
the generating function, we need to introduce some
definitions. First, similar to the vector space $\mathbb{C}\langle
X\rangle$ defined in the previous section, we define
$\mathbb{C}\langle \cL\rangle$ as the vector space generated by all
SVHPLs $\cL_\omega(z)\equiv\cL_\omega(-w)$. Note that there is a
natural linear map $\mathcal{P}_{\cL}:\mathbb{C}\langle X\rangle\to
\mathbb{C}\langle \cL\rangle$ sending a word $\omega$ to the SVHPL
$\cL_\omega(z)$. In order to incorporate the expansion parameter
$\eta$, we enlarge the vector space $\mathbb{C}\langle X\rangle$ to
the ring $\mathbb{C}\langle X\rangle[[\eta]]$ of formal power series
in $\eta$ with coefficients in $\mathbb{C}\langle X\rangle$. The ring
$\mathbb{C}\langle \cL\rangle[[\eta]]$ is defined in a similar
fashion, and the map $\mathcal{P}_{\cL}$ extends in an obvious manner to 
a map from $\mathbb{C}\langle X\rangle[[\eta]]$ to $\mathbb{C}\langle
\cL\rangle[[\eta]]$.

By observing patterns in the coefficients of the SVHPLs that appear
at low orders in the $\eta$ expansion
(see {\it e.g.}~eq.~\eqref{fcoeffcL}), 
and inspired by a very similar problem in $\cN=4$ super-Yang-Mills
theory~\cite{Pennington:2012zj} (see below), we have found an all-orders
formula that reproduces the first 10 orders of the expansion.
In order to describe this formula compactly,
we define the following elements of
$\mathbb{C}\langle X\rangle[[\eta]]$,
\beq\bsp
\mathcal{X}(x_0,x_1;\eta) &\,=\,
\left[1-\left(\frac{e^{-x_0 \eta}-1}{x_0}\right)x_1\right]^{-1}\,,\\
\cZ(x_0;\eta)&\,=\,\sum_{k=1}^{\infty}\left[
\sum_{n=0}^{k-1}(-x_0)^{k-n-1}\sum_{m=0}^{n} \, \frac{2^{m}}{(k-m)!} \,
\mathfrak{Z}(n,m)\right]\,\eta^k\, , \\
\label{eq:allorder}
\esp\eeq
where the $\mathfrak{Z}(n,m)$ are particular combinations of $\zeta$
values of uniform weight $n$. They are related to partial Bell
polynomials, and are generated by the series,
\beq\bsp
 \exp\left\{-\frac{x\,y}{2}\,
\left[2\gamma_E+\psi(1+x)+\psi(1-x)\right]\right\} 
&\,=\exp\left[y \sum_{k=1}^{\infty} \zeta_{2k+1} x^{2k+1}\right]\\ 
&\,= \sum_{n=0}^\infty \sum_{m=0}^\infty \mathfrak{Z}(n,m)\,x^n y^m\, .
\esp\eeq

We are now in the position to state our main conjecture: the BFKL
Green's function can be written, to all orders in the perturbative
expansion parameter $\eta$, as
\beq
\label{eq:LLAeq}
f(\vec{q}_{1\perp},\vec{q}_{2\perp},\Delta y) 
= \frac{1}{2}\delta^{(2)}(\vec{q}_{1\perp}-\vec{q}_{2\perp})
+ \frac{1}{2 \pi|\vec{q}_{1\perp}-\vec{q}_{2\perp}|^2}
\,\mathcal{P}_{\cL}\left[\mathcal{X}(x_0,x_1;\eta)\,\cZ(x_0;\eta)\right]\,.
\eeq
Equation~\eqref{eq:LLAeq} can be interpreted as a generating
function for the perturbative coefficients $f_k(w,w^\ast)$.  We
checked that our conjecture agrees with the integral
representation~\eqref{eq:fmc} up to 10 loops, by performing high
order series expansions of both sides around $|w|=0$.

We can separate out a power-law prefactor in $f_k$ by writing
\beq
f_k(w,w^\ast) = \frac{|w|}{|1+w|^2}\,F_k(w,w^\ast)\,,
\label{fkww}
\eeq
where the pure transcendental functions $F_k$ are given by,
\beq
F(w,w^\ast;\eta)
= \mathcal{P}_{\cL}\left[\mathcal{X}(x_0,x_1;\eta)\,\cZ(x_0;\eta)\right] 
= \sum_{k=1}^\infty F_k(w,w^\ast)\,\eta^k\,.
\label{FkXZ}
\eeq
We obtain for the first few loop orders,
\beq\bsp
F_1(w,w^\ast) &\,= 1\,,\\
F_2(w,w^\ast) &\,= - \cL_1 - \frac{1}{2} \cL_0 \,,\\
F_3(w,w^\ast) &\,= \cL_{1,1} 
+ \frac{1}{2} ( \cL_{0,1} + \cL_{1,0} ) + \frac{1}{6} \cL_{0,0} \,,\\
F_4(w,w^\ast) &\,= - \cL_{1,1,1} 
- \frac{1}{2} ( \cL_{0,1,1} + \cL_{1,0,1} + \cL_{1,1,0} )
- \frac{1}{4} \cL_{0,1,0} \\
&\hskip0.5cm\null
- \frac{1}{6} ( \cL_{0,0,1} + \cL_{1,0,0} )
- \frac{1}{24}  \cL_{0,0,0} + \frac{1}{3} \zeta_3 \,.
\label{fcoeffcL}
\esp\eeq
Note that in eq.~\eqref{fcoeffcL} we suppressed the dependence of
the functions on their arguments, {\it i.e.},
$\cL_{\omega}\equiv \cL_{\omega}(-w,-w^\ast)$. In practice, the form 
of the generating function~\eqref{FkXZ} was conjectured by explicitly evaluating the inverse Fourier-Mellin transform~\eqref{eq:fmc} for the first few values of $k$ in terms of single valued HPLs, and then conjecturing a generating function that reproduces not only the results for small $k$, but in addition also correctly predicts the functions $f_k$ for larger values of $k$. While for large values of $k$ an analytic evaluation of eq.~\eqref{eq:fmc} in terms of SVHPLs may become prohibitive, it is possible to evaluate the inverse Fourier-Mellin transform numerically.

We can write a few more loop orders if we first introduce some
more compact notation.  Since every word that appears is a binary
number, we can write it as a decimal number.  Because there can
be initial zeroes in the binary string, we keep track of the length of
the original word with a superscript in brackets.
For example, $\cL_{1,0,1}\to \ell_5^{[3]}$ and
$\cL_{0,1,0,1,1}\to \ell_{11}^{[5]}$.  In this notation, we have,
\beq\bsp
F_1 &\,= 1\,,\\
F_2 &\,= - \ell_{1}^{[1]} - \frac{1}{2} \ell_{0}^{[1]} \,,\\
F_3 &\,= \ell_{3}^{[2]} 
+ \frac{1}{2} ( \ell_{1}^{[2]} + \ell_{2}^{[2]} )
+ \frac{1}{6} \ell_{0}^{[2]} \,,\\
F_4 &\,= - \ell_{7}^{[3]}
- \frac{1}{2} ( \ell_{3}^{[3]} + \ell_{5}^{[3]} + \ell_{6}^{[3]} )
- \frac{1}{4} \ell_{2}^{[3]}
- \frac{1}{6} ( \ell_{1}^{[3]} + \ell_{4}^{[3]} )
- \frac{1}{24}  \ell_{0}^{[3]} + \frac{1}{3} \zeta_3 \,, \\
F_5 &\,= \ell_{15}^{[4]}
+ \frac{1}{2} ( \ell_{14}^{[4]} + \ell_{13}^{[4]} + \ell_{11}^{[4]} + \ell_{7}^{[4]} )
+ \frac{1}{6} ( \ell_{12}^{[4]} + \ell_{9}^{[4]} + \ell_{3}^{[4]} )
+ \frac{1}{4} ( \ell_{10}^{[4]} + \ell_{6}^{[4]} + \ell_{5}^{[4]} ) \\
&\hskip0.5cm\null
+ \frac{1}{24} ( \ell_{8}^{[4]} + \ell_{1}^{[4]} )
+ \frac{1}{12} ( \ell_{4}^{[4]} + \ell_{2}^{[4]} ) + \frac{1}{120} \ell_{0}^{[4]}
- \frac{1}{3} \zeta_3 
 \Bigl( \ell_{1}^{[1]} + \frac{1}{4} \ell_{0}^{[1]} \Bigr) \,, \\
F_6 &\,=  - \ell_{31}^{[5]}
- \frac{1}{2} ( \ell_{15}^{[5]} + \ell_{23}^{[5]} + \ell_{27}^{[5]} + \ell_{29}^{[5]}
    + \ell_{30}^{[5]} )
- \frac{1}{4} ( \ell_{11}^{[5]} + \ell_{13}^{[5]} + \ell_{14}^{[5]} + \ell_{21}^{[5]}
   + \ell_{22}^{[5]} + \ell_{26}^{[5]} ) \\
&\hskip0.5cm\null
- \frac{1}{6} ( \ell_{7}^{[5]} + \ell_{19}^{[5]} + \ell_{25}^{[5]} + \ell_{28}^{[5]} )
- \frac{1}{8} \ell_{10}^{[5]}
- \frac{1}{12} ( \ell_{5}^{[5]} + \ell_{6}^{[5]} + \ell_{9}^{[5]} 
   + \ell_{12}^{[5]} + \ell_{18}^{[5]} + \ell_{20}^{[5]} ) \\
&\hskip0.5cm\null
- \frac{1}{24} ( \ell_{3}^{[5]} + \ell_{17}^{[5]} + \ell_{24}^{[5]} )
- \frac{1}{36} \ell_{4}^{[5]}
- \frac{1}{48} ( \ell_{2}^{[5]} + \ell_{8}^{[5]} )
- \frac{1}{120} ( \ell_{1}^{[5]} + \ell_{16}^{[5]} )
- \frac{1}{720} \ell_{0}^{[5]} \\
&\hskip0.5cm\null
+ \frac{1}{3} \zeta_3 \Bigl( \ell_{3}^{[2]} + \frac{1}{2} \ell_{1}^{[2]} 
+ \frac{1}{4} \ell_{2}^{[2]} + \frac{1}{20} \ell_{0}^{[2]} \Bigr)
+ \frac{1}{60} \zeta_5 \,,\\
\label{fcoeffcLbig1}
\esp\eeq
\beq\bsp
F_7 &\,= \ell_{63}^{[6]}
+ \frac{1}{2} ( \ell_{31}^{[6]} + \ell_{47}^{[6]} + \ell_{55}^{[6]} 
+ \ell_{59}^{[6]} + \ell_{61}^{[6]} + \ell_{62}^{[6]} )
+ \frac{1}{4} ( \ell_{23}^{[6]} + \ell_{27}^{[6]} + \ell_{29}^{[6]} 
+ \ell_{30}^{[6]} \\
&\hskip0.5cm\null
    + \ell_{43}^{[6]} + \ell_{45}^{[6]} + \ell_{46}^{[6]}
    + \ell_{53}^{[6]} + \ell_{54}^{[6]} + \ell_{58}^{[6]} )
+ \frac{1}{6} ( \ell_{15}^{[6]} + \ell_{39}^{[6]} + \ell_{51}^{[6]} 
   + \ell_{57}^{[6]} + \ell_{60}^{[6]} ) \\
&\hskip0.5cm\null
+ \frac{1}{8} ( \ell_{21}^{[6]} + \ell_{22}^{[6]} + \ell_{26}^{[6]} + \ell_{42}^{[6]} )
+ \frac{1}{12} ( \ell_{11}^{[6]} + \ell_{13}^{[6]} + \ell_{14}^{[6]}
  + \ell_{19}^{[6]} + \ell_{25}^{[6]} + \ell_{28}^{[6]} + \ell_{37}^{[6]} 
  + \ell_{38}^{[6]} \\
&\hskip0.5cm\null
+ \ell_{41}^{[6]} + \ell_{44}^{[6]} + \ell_{50}^{[6]} + \ell_{52}^{[6]} )
+ \frac{1}{24} ( \ell_{7}^{[6]} + \ell_{10}^{[6]} + \ell_{18}^{[6]} 
+ \ell_{20}^{[6]} + \ell_{35}^{[6]} + \ell_{49}^{[6]} + \ell_{56}^{[6]} ) \\
&\hskip0.5cm\null
+ \frac{1}{36} ( \ell_{9}^{[6]} + \ell_{12}^{[6]} + \ell_{36}^{[6]} ) 
+ \frac{1}{48} ( \ell_{5}^{[6]} + \ell_{6}^{[6]} + \ell_{17}^{[6]} 
+ \ell_{24}^{[6]} + \ell_{34}^{[6]} + \ell_{40}^{[6]} ) \\
&\hskip0.5cm\null
+ \frac{1}{120} ( \ell_{3}^{[6]} + \ell_{33}^{[6]} + \ell_{48}^{[6]} )
+ \frac{1}{144} ( \ell_{4}^{[6]} + \ell_{8}^{[6]} )
+ \frac{1}{240} ( \ell_{2}^{[6]} + \ell_{16}^{[6]} )
+ \frac{1}{720} ( \ell_{1}^{[6]} + \ell_{32}^{[6]} ) \\
&\hskip0.5cm\null
+ \frac{1}{5040} \ell_{0}^{[6]} 
- \frac{1}{3} \zeta_3 \Bigl( \ell_{7}^{[3]} 
  + \frac{1}{2} ( \ell_{3}^{[3]} + \ell_{5}^{[3]} ) + \frac{1}{4} \ell_{6}^{[3]}
  + \frac{1}{6} \ell_{1}^{[3]} + \frac{1}{8} \ell_{2}^{[3]}
+ \frac{1}{20} \ell_{4}^{[3]} + \frac{1}{120} \ell_{0}^{[3]} \Bigr) \\
&\hskip0.5cm\null
- \frac{1}{60} \zeta_5 \Bigl( \ell_{1}^{[1]} + \frac{1}{6} \ell_{0}^{[1]} \Bigr)
+ \frac{1}{60} (\zeta_3)^2 \,.
\nonumber
\esp\eeq

In eq.~\eqref{fcoeffcLbig1} we have not yet made use
of the $\mathbb{Z}_2\times\mathbb{Z}_2$ symmetry
defined in eq.~\eqref{eq:Z2xZ2}.  Also, we have not applied any
shuffle identities in order to reduce the number of $\cL_{\omega}$
functions to a minimal set.  In the first step we
define~\cite{Dixon:2012yy} the projections onto definite
eigenstates under conjugation,\footnote{%
Note that we could also define eigenfunctions with
opposite parity under conjugation by
\beq
\overline{L}_\omega(z,\bar{z}) =\frac{1}{2}
\left[ \cL_\omega(z,\bar{z})
  +(-1)^{|\omega|}\cL_\omega(\bar{z},z)\right]\,.
\eeq
However, these functions are always products of the functions defined
in eq.~\eqref{eq:Lplusminus}~\cite{Dixon:2012yy}.}
\begin{equation}
\label{eq:Lplusminus}
L_\omega(z,\bar{z}) = \frac{1}{2}\left[ 
\cL_\omega(z,\bar{z}) - (-1)^{|\omega|}\,\cL_\omega(\bar{z},z)\right] \,,
\end{equation}
where $|\omega|$ is the weight or length of $\omega$.  Then
we construct the eigenstates with eigenvalues $\pm1$ under inversion,
\beq
L_{\omega}^\pm(z,\bar{z}) \equiv {1\over 2}
\,\left[L_\omega(z,\bar{z}) 
\pm L_\omega\Big({1\over z},{1\over \bar{z}}\Big)\right] \,.
\label{projinv}
\eeq
For a given word $\omega$, only the sign $(-1)^{|\omega|+d_\omega}$
in \eqn{projinv} leads to an irreducible function.
Here $d_\omega$ is the depth, or number of 1's in $\omega$, or the 
length of $\omega$ in the collapsed notation we employ below.
In this notation, repeated zeros in a word are
removed by letting $\vec{0}_{m-1}1\to m$, so for example
$L_{0,1,0,1,1}^+ \to L_{2,2,1}^+$. Finally, we can use shuffle identities,
based on eq.~\eqref{eq:cL_shuffle}, to express as many functions as
possible in terms of functions of lower weight (which are considered simpler).
It is known that a convenient basis of a shuffle algebra that enjoys this
property is given by the so-called \emph{Lyndon words}. A Lyndon word is a
word $w$ such that for every decomposition into two words $w = uv$, the
left word is lexicographically smaller than the right, $u < v$.
Every element of a shuffle algebra can be represented as a polynomial in
the Lyndon words.  The number of Lyndon words, and hence the number
of functions that are irreducible with respect to the shuffle identities,
are rather small at low weights.  At weight 1,2,3,4,5, there
are respectively only 2,1,2,3,6 such functions.  For the $L_\omega^\pm$
basis these functions are: $L_0^-, L_1^+; L_2^-; L_3^+, L_{2,1}^-;
L_4^-, L_{3,1}^+, L_{2,1,1}^-;
L_5^+, L_{4,1}^-, L_{3,2}^-, L_{3,1,1}^+, L_{2,2,1}^+, L_{2,1,1,1}^-$.

Applying this procedure to \eqn{fcoeffcLbig1}, we obtain,
\beq\bsp
F_1(w,w^\ast) &\,= 1\,,\\
F_2(w,w^\ast) &\,= -L_1^+\,,\\
F_3(w,w^\ast) &\,=\frac{1}{2}(L_1^+)^2-\frac{1}{24}\, (L_0^-)^2\,,\\
F_4(w,w^\ast) &\,=\frac{1}{6}L_3^+-\frac{1}{6}(L_1^+)^3+\frac{1}{3}\zeta_3\,,\\
F_5(w,w^\ast) &\,=-\frac{1}{6}L_3^+\,L_1^+-\frac{1}{12}L_{2,1}^-\,L_0^-
+\frac{1}{24}(L_2^-)^2
+\frac{1}{24}(L_1^+)^4 + \frac{1}{32}(L_0^-)^2\,(L_1^+)^2\\
&\, + \frac{11}{2880}(L_0^-)^4
 - \frac{1}{3}\zeta_3\,L_1^+\,,\\
F_6(w,w^\ast) &\,=-\frac{1}{10}L_5^+ -\frac{1}{3}L_{3,1,1}^+
-\frac{1}{6}L_{2,2,1}^+
+\frac{1}{12}L_3^+\,(L_1^+)^2+\frac{1}{144}L_3^+\,(L_0^-)^2\\
&\,+\frac{1}{12}L_{2,1}^-\,L_1^+\,L_0^-
-\frac{1}{120}(L_1^+)^5 -\frac{1}{48}(L_0^-)^2\,(L_1^+)^3
-\frac{1}{1152}(L_0^-)^4\,L_1^+\\
&\,+\frac{1}{6}\zeta_3\,(L_1^+)^2-\frac{3}{40}\zeta_3\,(L_0^-)^2
+\frac{1}{10}\zeta_5\,.
\label{fcoeffL}
\esp\eeq
Again in eq.~\eqref{fcoeffL} we have suppressed the dependence of
the functions on their arguments, {\it i.e.}, 
$L^\pm_{\omega}\equiv L^\pm_{\omega}(-w,-w^\ast)$.

Let us illustrate the procedure of passing from the $\cL_\omega$ functions
in eq.~\eqref{fcoeffcL} to the $L_\omega^\pm$ functions in eq.~\eqref{fcoeffL}
with the example of the functions $F_2$ and $F_3$.  Since
\beq\bsp
\label{eq:L0L1}
\cL_0(-w) &\,= \ln|w|^2 = L_0 = L_0^- \,, \\
\cL_1(-w) &\,= -\ln|1+w|^2 = L_1 = L_1^+ - \frac{1}{2} L_0^- \,,
\esp\eeq
we immediately obtain, using also eq.~\eqref{projinv},
\beq\bsp
F_2(w,w^*) &\,= \ln|1+w|^2 - \frac{1}{2}\ln|w|^2
  = \frac{1}{2}\ln|1+w|^2 + \frac{1}{2}\ln|1+1/w|^2 \\
  &\,= - \frac{1}{2} L_1(-w) - \frac{1}{2} L_1(-1/w) = -L_1^+\,.
\esp\eeq
For $F_3$, we can use the shuffle identities
\beq
\cL_{1,1} = \frac{1}{2}\cL_{1}^2{\rm~~and~~}\cL_{0,1} + \cL_{1,0} = \cL_0\,\cL_1
{\rm~~and~~}\cL_{0,0} = \frac{1}{2}\cL_{0}^2\,,
\eeq
and from eq.~\eqref{eq:L0L1} we obtain
\beq\bsp
F_3(w,w^*) &\,= \frac{1}{2}\ln^2|1+w|^2 - \frac{1}{2}\ln|1+w|^2\,\ln|w|^2
+\frac{1}{12} \ln^2|w|^2\\
&\,= \frac{1}{2} 
\biggl[ \frac{1}{2}\Bigl(  \ln|1+w|^2 + \ln|1+1/w|^2 \Bigr) \biggr]^2
- \frac{1}{24} \ln^2 |w|^2 \\
&\,=\frac{1}{2} (L_1^+)^2-\frac{1}{24}\, (L_0^-)^2\,.
\esp\eeq

Finally, we observe that \eqn{eq:LLAeq} is strikingly similar
to the corresponding formula describing the multi-Regge limit of the
six-point MHV and NMHV amplitudes in $\cN=4$ super-Yang-Mills
theory~\cite{Pennington:2012zj} in the leading-logarithmic approximation.
In fact, that formula inspired the form of \eqn{eq:LLAeq}.
The corresponding factors of $\mathcal{X}$ and $\mathcal{Z}$
are slightly different in the two cases.  There is an overall factor
of $x_1$ in the expressions for $\mathcal{Z}^{\rm MHV}$ and
$\mathcal{Z}^{\rm NMHV}$ in ref.~\cite{Pennington:2012zj}; this factor causes
the leading behavior of the LLA MHV and NMHV remainder functions in
the limit $|w|\to 0$ to be power suppressed. In the present case 
the $|w|\to 0$ behavior of the pure functions $F_k$ is power-unsuppressed.
(There is still power suppression coming from the rational prefactor $|w|$ in
\eqn{fkww}.)  The ordering of the $x_0$ and $x_1$ factors,
and the coefficients of the exponentials in $x_0\eta$, are slightly
different in the formula for $\mathcal{X}$ in the $\cN=4$ super-Yang-Mills
case, and there are also slightly different signs and factors of two
in the $\mathcal{Z}$ formulae.  Overall, however, the two types of formulae
bear a remarkably close resemblance.

The limiting behavior of $F_k(w,w^\ast)$ as $|w| \to 0$ is particularly
simple.  This limit corresponds to the region in which one tagging jet 
has much larger transverse momentum than the other.
If we neglect all terms that are suppressed by at least one
power of $|w|$, then we can drop all terms in \eqn{eq:allorder}
that contain a $x_1$.  In other words, we can set $\mathcal{X}\to1$.
We can also replace $x_0^p \to \cL_{\vec{0}_p} = (\ln |w|^2)^p /p!$, obtaining
from $\mathcal{Z}$,
\beq
F(w,w^\ast;\eta) = \sum_{k=1}^\infty \eta^k
\sum_{n=0}^{k-1} \frac{(-\ln|w|^2)^{k-n-1}}{(k-n-1)!} 
\sum_{m=0}^n \frac{2^m}{(k-m)!} \mathfrak{Z}(n,m)\ +\ {\cal O}(|w|)\,.
\label{Fksmallw}
\eeq
We can go further and resum this formula in the variable
$x = - \eta \ln|w|^2$, 
\beq
F(w,w^\ast;\eta) = 
\sum_{m=1}^\infty (2\eta)^m \sum_{j=1}^m \mathfrak{Z}(m-1,m-j)
 (2\sqrt{x})^{-j} I_{j}(2\sqrt{x})\ +\ {\cal O}(|w|)\,,
\label{resum_small_w}
\eeq
where the $I_j$ are modified Bessel functions.

\section{Analytic distributions for Mueller-Navelet jets}
\label{sec:mnjets}

In the previous section we have derived an all-orders expression for
the perturbative expansion of the LL BFKL Green's function. Using
eq.~\eqref{cross}, we can immediately write down the explicit
expression for the gluon-gluon cross section in the high-energy limit
to any loop order, in LL approximation.
In particular, for $k=1$ the dijet partonic cross
section (\ref{cross}) with the Green's function (\ref{eqn:greensvhpl})
becomes
\beq
\frac{d\hat\sigma^{(1)}_{gg}}{dp_{1\bot}^2dp_{2\bot}^2d\phi_{jj}} =
\frac{(C_A\as)^2}{4\pi p_{1\bot}^2 p_{2\bot}^2}\,
\frac{C_A\as\Delta y}{p_{1\bot}^2 + p_{2\bot}^2
 + 2 \sqrt{ p_{1\bot}^2\,p_{2\bot}^2 } \cos\phi_{jj}}\,,
\label{eq:oneloop}
\eeq
in agreement with ref.~\cite{DelDuca:1994ng}. Note that
eq.~\eqref{eq:oneloop} is divergent when $p_{1\bot}^2=p_{2\bot}^2$
and $\phi_{jj}=\pi$, {\it i.e.}~when the jets are back-to-back. 

While the results of the previous section are sufficient to obtain the
fully differential partonic dijet cross section in the high-energy
limit to any loop order, we show in the rest of this section that the
resulting expressions in terms of SVHPLs are particularly well suited
to performing the integration over the azimuthal angle and the
magnitude of the transverse momentum.  Thus we can obtain explicit
expressions for the dijet cross section in the high-energy limit at
leading logarithm that are inclusive in the transverse
momentum and exclusive in the azimuthal angle, or vice-versa, or inclusive in both.

\subsection{The azimuthal-angle distribution}
\label{sec:azim2}

The azimuthal-angle distribution is obtained by integrating the fully
differential cross section over the transverse momenta above a
threshold $E_\bot$. It admits the perturbative expansion,
\beq
\frac{d\hat\sigma_{gg}}{d\phi_{jj}} = \frac{\pi (C_A\as)^2}{2 E_\bot^2} 
\left[\delta(\phi_{jj}-\pi) 
+ \sum_{k=1}^\infty \frac{a_k(\phi_{jj})}{\pi}\, \eta^k \right]\,,
\label{eq:svazi}
\eeq
where the contribution of the $k^{\rm th}$ loop is given by the integral,
\beq
a_k(\phi_{jj}) = \frac{E^2_\bot}2\,  
\int_{E^2_\bot}^\infty 
\frac{dp_{1\bot}^2\,dp_{2\bot}^2}{(p_{1\bot}^2\,p_{2\bot}^2)^{3/2}} \,f_k(w,w^\ast)\,. 
\eeq
Changing variables to
\beq
\rho^2=|w|^2 = \frac{p_{1\bot}^2}{p_{2\bot}^2}
{\rm~~and~~} x = p_{1\bot}^2\,p_{2\bot}^2\,,
\eeq
the integration over $x$ becomes trivial, and we obtain
\beq\label{eq:ak_def}
a_k(\phi_{jj}) = \int_0^\infty \frac{d|w|}{|w|} f_k(w,w^\ast)
= 2\int_0^1 \frac{d\rho}{(1+\rho\,\varepsilon)(1+\rho\,\varepsilon^{-1})} 
F_k\left(\rho\,\varepsilon,\rho\,\varepsilon^{-1}\right) \,,
\eeq
with $\varepsilon=e^{-i\phi_{jj}}$, and where the last step follows from
eqs.~\eqref{eq:fk_symmetry} and \eqref{fkww}.

In the following we argue that the integral~\eqref{eq:ak_def} can
be evaluated easily if $F_k(w,w^\ast)$ is given in terms of
SVHPLs. From the definition of the SVHPLs, it is easy to see that we can
always write
\beq\label{eq:FkasHH}
F_k(w,w^\ast) = \sum_{i,j}c_{ij}\,H_{\omega_i}(-w)\,H_{\omega_j}(-w^\ast)\,,
\eeq
for some constants $c_{ij}$.
In order to perform the integration over the modulus of $w$, it is 
convenient to introduce a more general class of functions, namely the
so-called multiple polylogarithms defined as the iterated integrals,
\beq\label{eq:G_rec_def}
G(a_1,\ldots,a_n;z) = \int_0^z\frac{dt}{t-a_1}\,G(a_2,\ldots,a_n;t)\,,
\qquad a_i\in\mathbb{C}\,.
\eeq
Clearly, HPLs correspond to the special case of multiple
polylogarithms with all $a_i\in\{0,1\}$,
\beq\label{eq:HToG}
H_{a_1\ldots a_n}(z) = (-1)^p\,G(a_1,\ldots,a_n;z)\,,\qquad a_i\in\{0,1\}\,,
\eeq
where $p=\#\{a_i=1\}$. Multiple polylogarithms fulfill many identities
among themselves. In particular, they form a shuffle algebra (similar
to HPLs) and they satisfy the relation
\beq
G(k\,a_1,\ldots,k\,a_n;k\,z) 
= G(a_1,\ldots,a_n;z)\,,\quad\textrm{if }a_n,k\neq0\,.
\eeq
Using eq.~\eqref{eq:HToG} and this identity, $F_k$ in eq.~\eqref{eq:FkasHH}
may be written as
\beq\label{eq:G(|w|)}
F_k(\rho\varepsilon,\rho\varepsilon^{-1}) 
= \sum_{i,j}(-1)^{p_i+p_j}\,c_{ij}\,G(-\varepsilon^{-1}\,\omega_i;\rho)
\,G(-\varepsilon\,\omega_j;\rho)\,,
\eeq
where $p_i$ and $p_j$ denote the number of non-zero letters inside
$\omega_i$ and $\omega_j$, {\it cf.}~eq.~\eqref{eq:HToG}.
The indices of the multiple polylogarithms appearing inside are obtained
by multiplying every letter (0 or 1) in the word $\omega_i$
($\omega_j$) by $-\varepsilon^{-1}$ ($-\varepsilon$).
Inserting eq.~\eqref{eq:G(|w|)} into eq.~\eqref{eq:ak_def}, performing
a partial fraction decomposition of 
$1/[(1+\rho\,\varepsilon)(1+\rho\,\varepsilon^{-1})]$, and using
the shuffle algebra properties of multiple polylogarithms, we see that
the integration over $\rho$ can easily be performed using the recursive
definition~\eqref{eq:G_rec_def}. As a result, we can write
$a_k(\phi_{jj})$ as a linear combination of multiple polylogarithms
$G(a_1,\ldots,a_n;1)$ with
$a_i\in\{0,-\varepsilon,-\varepsilon^{-1}\}$.

It turns out that the results for the functions $a_k(\phi_{jj})$ can be 
recast in a form that only involves harmonic polylogarithms. Indeed,
multiple polylogarithms satisfy various intricate identities, and recently
a lot of progress was made in simplifying complicated expressions by
using the so-called symbol of a transcendental
function~\cite{symbolsC,symbolsB,symbols,Goncharov:2010jf,Duhr:2011zq}.
In particular, the symbol of $a_k(\phi_{jj})$ can always be written such
that all its entries are drawn
from the set $\{\varepsilon^2,1-\varepsilon^2\}$, which are symbols of
HPLs with arguments $\varepsilon^2=e^{-2i\phi_{jj}}$.  Defining
$H_{i,j,\ldots}\equiv H_{i,j,\ldots}(\varepsilon^2)$, we find explicitly
up to six loops,
%
%
%
%
\beq\label{eq:ak_result}
a_k(\phi_{jj}) = \frac{\mathrm{Im}\,A_k(\phi_{jj})}{\sin\phi_{jj}}\,,
\eeq
with
\beq\bsp\label{A_k}
A_1(\phi_{jj}) &\,= -\frac{1}{2} H_0\,,\\
A_2(\phi_{jj}) &\, = H_{1,0}\,,\\
A_3(\phi_{jj}) &\,= \frac{2}{3} H_{0,0,0} - 2 H_{1,1,0} + \frac{5}{3} \zeta_2 H_0
- i \pi \, \zeta_2 \,,\\
A_4(\phi_{jj}) &\,= -\frac{4}{3} H_{0,0,1,0} - H_{0,1,0,0} - \frac{4}{3} H_{1,0,0,0}
+ 4 H_{1,1,1,0}
- \zeta_2 \biggl( 2 H_{0,1} + \frac{10}{3} H_{1,0} \biggr)\\
&\hskip0.5cm\null
+ \frac{4}{3} \zeta _3 \,H_0
+i \pi  \Big( 2 \zeta_2 H_{1} - 2 \zeta _3 \Big)\,,\\
A_5(\phi_{jj}) &\, = -\frac{46}{15} H_{0,0,0,0,0}
+ \frac{8}{3} H_{0,0,1,1,0} + 2 H_{0,1,0,1,0}
+ 2 H_{0,1,1,0,0} + \frac{8}{3} H_{1,0,0,1,0} + 2 H_{1,0,1,0,0}
\\
&\hskip0.5cm\null
+ \frac{8}{3} H_{1,1,0,0,0} - 8 H_{1,1,1,1,0}
- \zeta_2 \biggl( \frac{33}{5} H_{0,0,0} - 4 H_{0,1,1} - 4 H_{1,0,1}
    - \frac{20}{3} H_{1,1,0} \biggr)\\
&\hskip0.5cm\null
- \zeta_3 \biggl( 2 H_{0,1} + \frac{8}{3} H_{1,0} \biggr)
+ \frac{217}{15} \zeta_4 H_0 \\
&\hskip0.5cm\null
+i \pi  \biggl[ \zeta_2 \biggl( \frac{10}{3} H_{0,0} - 4 H_{1,1} \biggr)
 + 4 \zeta_3 H_1 - \frac{10}{3} \zeta_4 \biggr]\,,
\esp\eeq
%
%
\beq\bsp\nonumber
A_6(\phi_{jj}) &\, =  \frac{92}{15} H_{0,0,0,0,1,0} + \frac{17}{3} H_{0,0,0,1,0,0}
+ \frac{52}{9} H_{0,0,1,0,0,0} - \frac{16}{3} H_{0,0,1,1,1,0} \\
&\hskip0.5cm\null
+ \frac{17}{3} H_{0,1,0,0,0,0}
- 4 H_{0,1,0,1,1,0} - 4 H_{0,1,1,0,1,0}
- 4 H_{0,1,1,1,0,0} + \frac{92}{15} H_{1,0,0,0,0,0} \\
&\hskip0.5cm\null
- \frac{16}{3} H_{1,0,0,1,1,0} - 4 H_{1,0,1,0,1,0} - 4 H_{1,0,1,1,0,0}
- \frac{16}{3} H_{1,1,0,0,1,0} - 4 H_{1,1,0,1,0,0} \\
&\hskip0.5cm\null
- \frac{16}{3} H_{1,1,1,0,0,0}
+ 16 H_{1,1,1,1,1,0}
- \zeta_2 \biggl( - \frac{34}{3} H_{0,0,0,1} - \frac{112}{9} H_{0,0,1,0}
  - 12 H_{0,1,0,0} \\
&\hskip0.5cm\null
+ 8 H_{0,1,1,1} - \frac{66}{5} H_{1,0,0,0} + 8 H_{1,0,1,1}
  + 8 H_{1,1,0,1} + \frac{40}{3} H_{1,1,1,0} \biggr) \\
&\hskip0.5cm\null
- \zeta_3 \biggl( \frac{92}{15} H_{0,0,0} - 4 H_{0,1,1} - 4 H_{1,0,1}
          - \frac{16}{3} H_{1,1,0} \biggr)
- \zeta_4 \biggl( \frac{77}{3} H_{0,1} + \frac{434}{15} H_{1,0} \biggr) \\
&\hskip0.5cm\null
- \biggl( \frac{4}{5} \zeta_5 - \frac{262}{15} \zeta_2 \zeta_3 \biggr) H_{0}
+ i\pi \biggl[ \zeta_2 \biggl( -\frac{20}{3} H_{0,0,1} - 6 H_{0,1,0}
              - \frac{20}{3} H_{1,0,0} + 8 H_{1,1,1} \biggr) \\
&\hskip1.5cm\null
              + \zeta_3 \biggl( \frac{20}{3} H_{0,0} - 8 H_{1,1} \biggr)
        + \frac{20}{3} \zeta_4 H_{1} - \frac{8}{3} \zeta_2 \zeta_3
       \biggr] \,.
\esp\eeq
Similar results can be obtained at higher loop orders in exactly the
same fashion.

A few comments are in order about the behavior of $a_k(\phi_{jj})$
in eq.~\eqref{eq:ak_result}, in the limits $\phi_{jj} \to 0$ and
$\phi_{jj} \to \pi$, and at the value $\phi_{jj} = \pi/2$.

The limit $\phi_{jj} \to 0$ should be nonsingular
in perturbation theory, since the configuration in which both tagging
jets are at the same azimuthal angle requires a large amount of
additional transverse momentum radiated from the ladder.
Naively, eq.~\eqref{eq:ak_result} would appear to diverge as
$\phi_{jj} \to 0$, from the factor of $\sin\phi_{jj}$ in the denominator.
However, the numerator factor $A_k(\phi_{jj})$ also vanishes linearly
with $\phi_{jj}$, resulting in the following finite values:
\beq\bsp
a_1(0) &\,= 1 \,,\\
a_2(0) &= 2 \,,\\
a_3(0) &= 4 - \frac{4}{3} \, \zeta_2 \,,\\
a_4(0) &= 8 - \frac{8}{3} \, \zeta_2 - 2 \, \zeta_3 \,,\\
a_5(0) &= 16 - \frac{16}{3} \, \zeta_2 - 4 \, \zeta_3 
             - \frac{8}{5} \, \zeta_4 \,,\\
a_6(0) &= 32 - \frac{32}{3} \, \zeta_2 - 8 \, \zeta_3 
       - \frac{16}{5} \, \zeta_4
       - \frac{58}{9} \, \zeta_5 + \frac{128}{45} \, \zeta_2 \zeta_3 \,.
\label{a_k_phi_to_0}
\esp\eeq

The case of jets at right angles, $\phi_{jj} = \pi/2$, is also nonsingular
and can be given analytically in terms of simple constants for
low loop orders, using \eqn{A_k}.  We find,
\beq\bsp
a_1(\pi/2) &\,= \frac{\pi}{2} \,,\\
a_2(\pi/2) &= \pi \, \ln 2 \,,\\
a_3(\pi/2) &= \pi \, \ln^2 2 \,,\\
a_4(\pi/2) &= \pi \biggl[ \frac{2}{3} \, \ln^3 2
         + \frac{1}{6} \, \zeta_3 \biggr] \,,\\
a_5(\pi/2) &= \pi \biggl[  \frac{1}{3}\,\ln^42+ \frac{1}{3}\,\zeta_3\,\ln2-\frac{1}{8}\,\zeta_4 \biggr] \,,\\
a_6(\pi/2) &= \pi \biggl[ \frac{2 }{15}\,\ln ^52+\frac{1}{3} \zeta_3 \,\ln ^22-\frac{1}{4}\, \zeta_4\, \ln 2+\frac{7 }{60}\,\zeta_2\, \zeta_3-\frac{19 }{120}\,\zeta_5 \biggr] \,.
\label{a_k_phi_pi_over_2}
\esp\eeq

The limit $\phi_{jj}\to\pi$ behaves differently because it has
a $\delta$-function singularity at Born level.  The fixed-order
expansions should be singular in this region, even in the LL approximation.
On the other hand, the BFKL resummation can cure the fixed-order divergence,
as often happens in many other contexts.
%
%
We can see this explicitly by analyzing the behavior of $F_k(w,w^\ast)$
in eq.~\eqref{fcoeffL} as $w\to -1$, corresponding to
the Born configuration with $p_{1\bot}^2 = p_{2\bot}^2$ and $\phi_{jj}=\pi$.
For this purpose we need the values of SVHPLs at $w=-1$ ($z=1$).
These values have been studied in a recent paper~\cite{Brown:2013gia}.
In general, they are either zero or multiple zeta values.
Using the Lyndon basis for $L_\omega^\pm$, the only divergent basis
function is $L_1^+ = -\ln|1+w|^2 + \tfrac{1}{2} \ln|w|^2 \approx -\ln|1+w|^2$.
Then, from \eqn{fcoeffL} we see that the dominant behavior of
$F_k$ as $w\rightarrow-1$ is
\beq
\lim_{w\to -1} F_k(w,w^\ast) = (-1)^{k-1}\, \frac{(L_1^+)^{k-1}}{(k-1)!} + \ldots
= \frac{\left(\ln |1+w|^2\right)^{k-1}}{(k-1)!} + \ldots\,,
\eeq
where we neglected terms that are subleading as $w\to -1$.
Then the most-singular contribution to the coefficient $a_k(\phi_{jj})$
at the $k^{\rm th}$ loop order becomes,
\beq
\lim_{\phi_{jj}\to\pi} a_k(\phi_{jj}) = 2 \int_0^1 \frac{d|w|}{|1+w|^2} 
\frac{\left(\ln |1+w|^2\right)^{k-1}}{(k-1)!} + \ldots\,.
\label{eq:limak}
\eeq
Once we insert \eqn{eq:limak} into the azimuthal-angle
distribution (\ref{eq:svazi}), we can resum the loop expansion
for the most singular behavior.  We obtain,
\beq
\frac{d\hat\sigma_{gg}}{d\phi_{jj}} = \frac{\pi (C_A\as)^2}{2 E_\bot^2} 
\left[\delta(\phi_{jj}-\pi) + \frac2{\pi}\, \frac{C_A\as}{\pi}
\, \Delta y\, \int_0^1 d|w| \left(|1+w|^2\right)^{-1+ \eta}
\right]\,,
\eeq
where $w=|w|e^{-i\phi_{jj}}$.  The would-be divergence in the azimuthal-angle
distribution for $\phi_{jj}=\pi$, from $w=-1$ in the integral, has been
regulated by the factor of $\eta$ in the exponent.
This is in agreement with the finiteness of the resummed saddle-point
expression in Section~\ref{sec:saddlept}.

\subsection{The transverse-momentum distribution}
\label{sec:tram}

In this section we show how one can compute in a similar way the
transverse-momentum distribution, obtained by integrating the fully
differential cross section over the azimuthal angle. It admits the
perturbative expansion
\beq
\frac{d\hat\sigma_{gg}}{dp_{1\bot}^2 dp_{2\bot}^2}
= \frac{\pi (C_A\as)^2}{2 p_{1\bot}^2 p_{2\bot}^2}\, 
\left[ \delta(p_{1\bot}^2-p_{2\bot}^2) 
+ \frac1{2\pi\, \sqrt{ p_{1\bot}^2\,p_{2\bot}^2 } }\,
b(\rho;\eta)\right]\,,
\label{eq:transvdis}
\eeq
where we used the same parametrization as for the azimuthal-angle
distribution (with $\phi=-\phi_{jj}$ here),
\beq 
w=\rho\,\varepsilon\,, 
\qquad \rho = |w| = \sqrt{\frac{p_{1\bot}^2}{p_{2\bot}^2}}
 \,, \qquad \varepsilon= e^{i\phi} \,.
\label{eq:rho}
\eeq
The function $b(\rho;\eta)$ admits the perturbative expansion
\beq
b(\rho;\eta) = \sum_{k=1}^\infty b_k(\rho)\, \eta^k\,,
\eeq
and the contribution of the $k^{\rm th}$ loop is
\beq\bsp\label{eq:bk_integral}
b_k(\rho) &\,= \int_{-\pi}^\pi d\phi \, f_k(w,w^\ast)\\
&\,= 2\int_{0}^\pi d\phi \, f_k(w,w^\ast)\\
&\,= -2\,i\,\int_{\mathcal{C}}\frac{{d}\varepsilon}{\varepsilon}
\,f_k\left(\rho\,\varepsilon,{\rho}\,{\varepsilon^{-1}}\right)\\
&\,=-2\,i\,\rho\,\int_{\mathcal{C}}
\frac{d\varepsilon}{(\rho+\varepsilon)(1+\rho\,\varepsilon)}
\,F_k\left(\rho\,\varepsilon,{\rho}\,{\varepsilon^{-1}}\right)\,,
\esp\eeq
where the second equality follows from the fact that $f_k(w,w^\ast)$ is real,
{\it i.e.}, an even function of $\phi$. The integration contour is given by
\beq
\mathcal{C} = \{\varepsilon\in \mathbb{C}|\,|\varepsilon|
=1 \textrm{ and Im}(\varepsilon)>0\}\,.
\eeq
In the following we use the symmetry under inversion of $w$ to let
$0<\rho<1$.  The integrand~\eqref{eq:bk_integral} has singularities at
$\varepsilon=-\rho$ and $\varepsilon=-1/\rho$. Our goal is deform the
contour $\mathcal{C}$ to the straight line $[-1,1]$. Then, after the contour 
deformation the singularity at
$\varepsilon=-1/\rho$ lies outside the integration region, but the
singularity $\varepsilon=-\rho$ does not.  The correct way to avoid
the singularity is to assign a small positive imaginary part to $\rho$,
\beq
\rho\to\rho+i0\,.
\eeq
In particular, we need the identity
\beq
\ln(\rho+i0-1) = \ln(1-\rho) + i\pi\,.
\eeq
We can thus rewrite eq.~\eqref{eq:bk_integral} as
\beq
b_k(\rho) =-2\,i\,\rho\,\int_{-1}^1
\frac{d\varepsilon}{(\rho+i0+\varepsilon)(1+\rho\,\varepsilon)}
\,F_k\left(\rho\,\varepsilon,{\rho}\,{\varepsilon^{-1}}\right)\,,
\quad 0<\rho<1\,.
\eeq
The remaining integral is easily performed in terms of HPLs, by
following the same strategy outlined in Section~\ref{sec:azim2}.
Writing $H_{i,j,\ldots}\equiv H_{i,j,\ldots}(\rho^2)$\,, we obtain
explicitly for the first six loops,
\beq
b(\rho;\eta) = \frac{2\,\pi\,\rho}{1-\rho^2}\,B(\rho;\eta) = \frac{2\,\pi\,\rho}{1-\rho^2}\,\sum_{k=1}^\infty B_k(\rho)\, \eta^k\,,
\eeq
with
\beq\bsp\label{eq:Bk_results}
B_1(\rho) &\,= 1\,,\\
B_2(\rho) &\,= - \frac{1}{2}\,H_0 - 2 H_1 \,,\\
B_3(\rho)&\,=\frac{1}{6}H_{0,0}+2 H_{0,1}+H_{1,0}+4 H_{1,1}\,,\\
B_4(\rho)&\,=-\frac{1}{24} H_{0,0,0}-\frac{4}{3} H_{0,0,1}-H_{0,1,0}-4 H_{0,1,1}
-\frac{1}{3} H_{1,0,0}-4 H_{1,0,1}-2 H_{1,1,0}\\
&\hskip0.5cm\null
-8 H_{1,1,1}+\frac{1}{3}\,\zeta _3\,,\\
B_5(\rho)&\,=\frac{1}{120} H_{0,0,0,0}+\frac{2}{3} H_{0,0,0,1}
+\frac{2}{3} H_{0,0,1,0}+\frac{8}{3} H_{0,0,1,1}+\frac{1}{3} H_{0,1,0,0}
+4 H_{0,1,0,1}\\
&\hskip0.5cm\null
+2 H_{0,1,1,0}+8 H_{0,1,1,1}+\frac{1}{12} H_{1,0,0,0}+\frac{8}{3} H_{1,0,0,1}
+2 H_{1,0,1,0}+8 H_{1,0,1,1}\\
&\hskip0.5cm\null
+\frac{2}{3} H_{1,1,0,0}+8 H_{1,1,0,1}+4 H_{1,1,1,0}+16 H_{1,1,1,1}
+ \zeta_3 \biggl( - \frac{1}{12} H_0 - \frac{2}{3} H_1 \biggr)\,,\\
B_6(\rho)&\,= -\frac{1}{720} H_{0,0,0,0,0}-\frac{4}{15} H_{0,0,0,0,1}
-\frac{1}{3} H_{0,0,0,1,0}-\frac{4}{3} H_{0,0,0,1,1}-\frac{2}{9} H_{0,0,1,0,0}\\
&\hskip0.5cm\null
-\frac{8}{3} H_{0,0,1,0,1}-\frac{4}{3} H_{0,0,1,1,0}
-\frac{16}{3} H_{0,0,1,1,1}-\frac{1}{12} H_{0,1,0,0,0}-\frac{8}{3} H_{0,1,0,0,1}\\
&\hskip0.5cm\null
-2 H_{0,1,0,1,0}-8 H_{0,1,0,1,1}-\frac{2}{3} H_{0,1,1,0,0}-8 H_{0,1,1,0,1}
-4 H_{0,1,1,1,0}\\
&\hskip0.5cm\null
-16 H_{0,1,1,1,1}-\frac{1}{60} H_{1,0,0,0,0}
-\frac{4}{3} H_{1,0,0,0,1}-\frac{4}{3} H_{1,0,0,1,0}-\frac{16}{3} H_{1,0,0,1,1}\\
&\hskip0.5cm\null
-\frac{2}{3} H_{1,0,1,0,0}-8 H_{1,0,1,0,1}
-4 H_{1,0,1,1,0}-16 H_{1,0,1,1,1}-\frac{1}{6} H_{1,1,0,0,0}\\
&\hskip0.5cm\null
-\frac{16}{3} H_{1,1,0,0,1}-4 H_{1,1,0,1,0}
-16 H_{1,1,0,1,1}-\frac{4}{3} H_{1,1,1,0,0}-16 H_{1,1,1,0,1}\\
&\hskip0.5cm\null
-8 H_{1,1,1,1,0} -32 H_{1,1,1,1,1}
+ \zeta_3 \biggl( \frac{1}{60} H_{0,0} + \frac{2}{3} H_{0,1}
            + \frac{1}{6} H_{1,0} + \frac{4}{3} H_{1,1} \biggr)\\
&\hskip0.5cm\null
+\frac{1}{60}\zeta _5\,.
\esp\eeq
\begin{center}
\begin{figure}[!t]
\begin{center}
\includegraphics[scale=1.6]{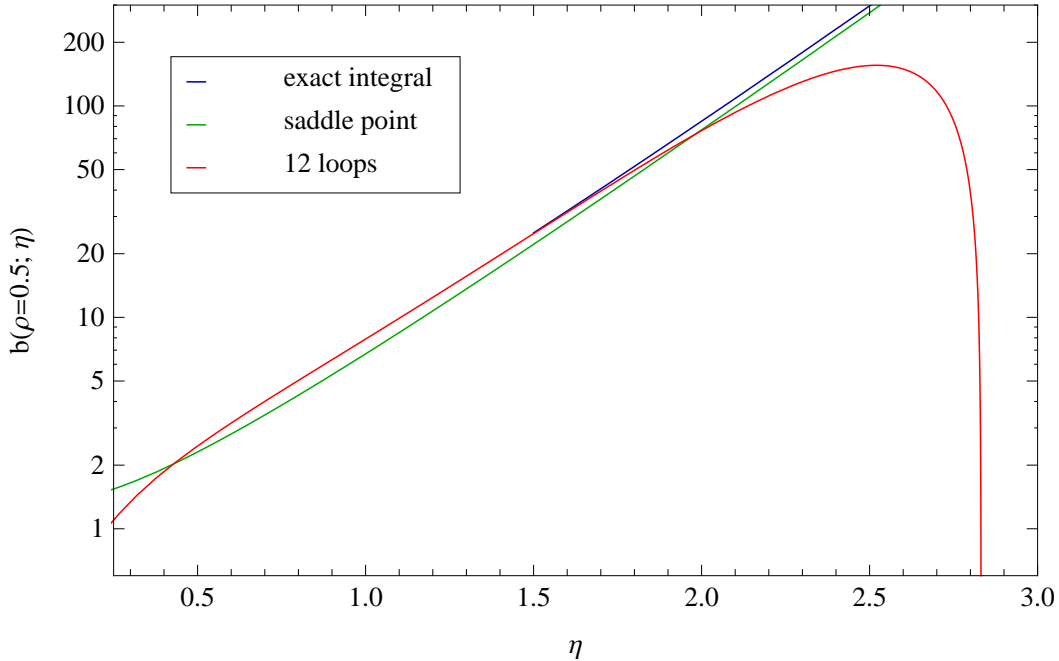}
\caption{\label{fig:Bk_plot_0.5}The transverse momentum distribution
$b(\rho;\eta)$ in the high-energy limit as a function of the rapidity
for $\rho = 0.5$ evaluated perturbatively through twelve loops (red), 
and compared to the exact Mellin integral (blue)
and its saddle-point approximation (green).}
\end{center}
\end{figure}
\end{center}

\begin{center}
\begin{figure}[!t]
\begin{center}
\includegraphics[scale=1.6]{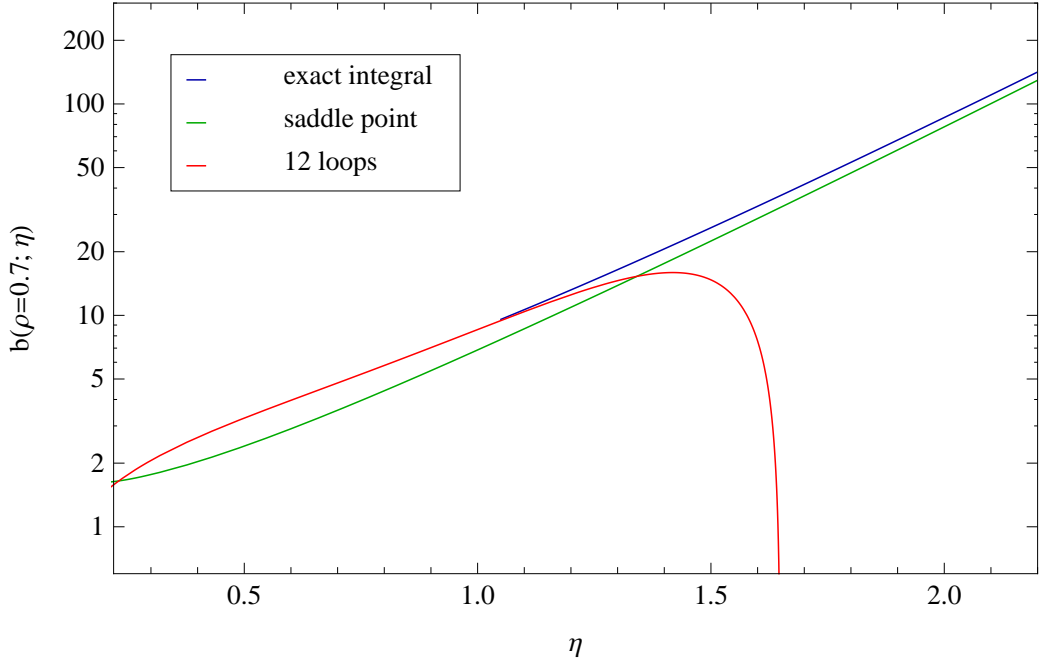}
\caption{\label{fig:Bk_plot_0.7}Same as Fig.~\ref{fig:Bk_plot_0.5},
but for $\rho=0.7$.}
\end{center}
\end{figure}
\end{center}

Remarkably, these formulas for $B_k(\rho)$ follow from essentially
the same generating function~\eqref{FkXZ} that we found for
$F_k(w,w^\ast)$:
\beq
B(\rho;\eta) = \mathcal{P}_{H}\left[\mathcal{X}(x_0,x_1;2\,\eta)\,\cZ(x_0;\eta)\right]\,.
\label{BkXZ}
\eeq
The only two differences with respect to \eqn{FkXZ} are that
$2\eta$, not $\eta$, appears as the argument of $\mathcal{X}$, and the
map $\mathcal{P}_{H}$ sends a word $\omega$ to the HPL
$H_\omega(\rho^2)$ instead of to the SVHPL $\cL_\omega(-w,-w^\ast)$.
We have not proven this result to all orders, but we have checked that it 
reproduces the analytical results for the coefficients $B_k(\rho)$ through 
six loops shown in eq.~\eqref{eq:Bk_results}. 
In addition, we have checked eq.~\eqref{BkXZ} up to twelve loops
numerically, by performing the integral in the Mellin representation of
$b(\rho,\eta)$,
\beq\label{eq:b_exact}
b(\rho;\eta) = \int_{-\infty}^{+\infty}d\nu\,\rho^{2i\nu}\,e^{\eta\,\chi_{\nu,0}}
\simeq \sqrt{\frac{\pi}{14\,\zeta_3\,\eta}}
\exp\left[4\eta\,\ln2-\frac{\ln^2\rho^2}{56\,\zeta_3\,\eta}\right] \,,
\eeq
obtained from eqs.~\eqref{green} and~\eqref{saddle} by integrating the
azimuthal angle over the range $[0,2\pi]$. More precisely, we have computed
the Mellin integral order-by-order in $\eta$ by closing the integration
contour in the upper half plane and numerically summing up the residues. 
Finally, we compare the transverse momentum distribution $b(\rho;\eta)$
truncated at twelve loops, as obtained from the generating
functional~\eqref{BkXZ}, with the exact Mellin integral
and its saddle-point approximation given in eq.~\eqref{eq:b_exact}.
We perform the comparison as a function of $\eta$ ({\it i.e.}~the rapidity)
for two selected values of $\rho$ in Fig.~\ref{fig:Bk_plot_0.5}
and~\ref{fig:Bk_plot_0.7}.  We observe that there is a very good
agreement between the exact integral and its perturbative expansion
over a wide range of rapidities. 

\subsection{The total cross section}
\label{sec:totc}

We can now perform a final integration
over the transverse momentum variable $\rho$.
The coefficients in the perturbative expansion contain a logarithmic
divergence as $\rho\rightarrow1$, so we will cut off the integral
at $\rho^2=1-\de$, after mapping the integral to the region
$p_{1\bot}^2 < p_{2\bot}^2$, or $0 < \rho < 1$.
We define the regulated total cross section,
\beq\bsp\label{totdelta1}
\hat\sigma_{gg}(\de) &\,\equiv 2 \int_{E_\bot^2}^\infty dp_{1\bot}^2 
\int_{p_{1\bot}^2/(1-\de)}^\infty dp_{2\bot}^2 
 \frac{d\hat\sigma_{gg}}{dp_{1\bot}^2 dp_{2\bot}^2}\\
&\,= 2 \int_{E_\bot^2}^\infty dp_{1\bot}^2 \, p_{1\bot}^2 
\int_0^{1-\de} \frac{d\rho^2}{\rho^4}
\frac{d\hat\sigma_{gg}}{dp_{1\bot}^2 dp_{2\bot}^2} \,.
\esp\eeq
Using eq.~\eqref{eq:transvdis}, we have
\beq\bsp\label{totdelta2}
\hat\sigma_{gg}(\de) &\,= (C_A\as)^2 \int_{E_\bot^2}^\infty 
\frac{dp_{1\bot}^2}{(p_{1\bot}^2)^2}
  \int_0^{1-\de} d\rho^2 \, \frac{b(\rho)}{2\rho}\\
&\,= \frac{\pi (C_A\as)^2}{E_\bot^2} \sum_{k=1}^\infty \eta^k 
\int_0^{1-\de} d\rho^2 \frac{B_k(\rho)}{1-\rho^2} \,.
\esp\eeq
Comparing with eq.~\eqref{eq:mnexp}, we see that
\beq
I_k(\de) \equiv 2 \int_0^{1-\de} d\rho^2 \, \frac{B_k(\rho)}{1-\rho^2} 
\label{Ikdef}
\eeq
should correspond to the coefficients $f_{0,k}$ in the
Mueller-Navelet dijet cross section~(\ref{eq:mnexp}). However,
at fixed order $k$ there will be a logarithmic divergence as
$\delta\rightarrow0$, for the reason discussed at the end of
Section~\ref{sec:azim2}.  As we did there, we will have to perform
the sum over $k$ for the divergent terms, which will regulate the divergence.
Then we can take the limit $\delta\rightarrow0$.
Whereas the discussion in Section~\ref{sec:azim2} concerned only
the leading logarithms at a given order in the $\eta$ expansion,
here we will provide a (conjectured) expression accurate to
all logarithmic orders.

Because the $B_k(\rho)$ are linear combinations of harmonic polylogarithms
with argument $\rho^2$, the integral can be performed for any 
value of $\de$, according to eq.~\eqref{eq:Hintdef}, by prepending a ``1'' to
each HPL weight vector and setting the argument to $1-\de$.
We then expand the result for small $\de$, neglecting power-suppressed
terms in $\de$.  We find the following result through nine loops,
\beq
I_k(\de) = f_{0,k}
- \sum_{m=0}^k Z_m \frac{(2\ln\de)^{k-m}}{(k-m)!} + {\cal O}(\de),
\label{Ikresult}
\eeq
where $f_{0,k}$ are the Mueller-Navelet coefficients~(\ref{MNtotalcross})
and the $Z_m$ coefficients are given through seven loops by,
\beq\bsp
\label{Zcoeffs}
Z_0 &\,= 1 \,, \\
Z_1 &\,= 0 \,, \\
Z_2 &\,= \zeta_2 \,, \\
Z_3 &\,= \frac{8}{3} \zeta_3 \,, \\
Z_4 &\,= \frac{19}{4} \zeta_4 \,, \\
Z_5 &\,= \frac{32}{5} \zeta_5 + \frac{8}{3} \zeta_2\, \zeta_3 \,, \\
Z_6 &\,= \frac{275}{16} \zeta_6 + \frac{32}{9} \zeta_3^2 \,, \\
Z_7 &\,= \frac{128}{7} \zeta_7 + \frac{38}{3} \zeta_3\, \zeta_4
   + \frac{32}{5} \zeta_2\, \zeta_5 \,.
\esp\eeq
These coefficients are consistent with the following all-orders
expression:
\beq
\label{Z_eta}
Z(\eta) \equiv \sum_{m=0}^\infty Z_m \eta^m
= \exp\biggl[ \, \sum_{k=2}^\infty \eta^k 
          \, \Bigl( 2^k - 1 - (-1)^k \Bigr) \, \frac{\zeta_k}{k} \biggr] =e^{-2\gamma_E\eta}\,\frac{\Gamma(1-2\eta)}{\Gamma(1-\eta)\,\Gamma(1+\eta)}\,,
\eeq
where $\gamma_E=-\Gamma'(1)$ denotes the Euler-Mascheroni constant. We explicitly checked that the all orders expression reproduces the correct coefficients $Z_m$ through 13 loops.

While each perturbative coefficient is logarithmically divergent,
when we resum the series the divergence goes away:
\beq\bsp
\label{fixIde}
I(\de) \equiv \sum_{k=0}^\infty I_k(\de) \eta^k
&\,= \sum_{k=0}^\infty f_{0,k} \eta^k
   - \sum_{m=0}^\infty Z_m \eta^m \sum_{l=0}^\infty \frac{(2\eta\ln\de)^l}{l!}
   + {\cal O}(\de)\\
&\,= \sum_{k=0}^\infty f_{0,k} \eta^k - Z(\eta) \de^{2\eta} + {\cal O}(\de)\,.
\esp\eeq
We now take the limit $\de\rightarrow0$ in the last form of this equation.
This limit allows us to identify the coefficients $f_{0,k}$ in 
eq.~\eqref{Ikresult} with the Mueller-Navelet coefficients
defined in eq.~\eqref{eq:mnexp}.

Finally, the next few Mueller-Navelet coefficients can be determined
analytically in this way,
\beq\bsp
\label{MNtotalcrossmore}
f_{0,6} &\,= \frac{13}{4}\,\zeta_3^2+\frac{3737}{120}\,\zeta_6\,,\\
f_{0,7} &\,= -\frac{87}{5}\,\zeta_3\,\zeta_4-\frac{116}{9}\,\zeta_2\,\zeta_5-\frac{3983}{144}\,\zeta_7\,,  \\
f_{0,8} &\,= -\frac{37}{75}\,\zeta_{5,3}+\frac{64}{15}\,\zeta_2\,\zeta_3^2+\frac{369}{20}\,\zeta_5\,\zeta_3+\frac{50606057}{453600}\,\zeta_8\,, \\
f_{0,9} &\,= -\frac{139}{60}\,\zeta_3^3-\frac{15517}{252}\,\zeta_6\,\zeta_3-\frac{3533}{63}\,\zeta_4\,\zeta_5-\frac{557}{15}\,\zeta_2\,\zeta_7-\frac{5215361}{60480}\,\zeta_9\,, \\
f_{0,10} &\,= -\frac{2488}{4725}\,\zeta_{5,3}\,\zeta_2-\frac{94721}{211680}\,\zeta_{7,3}+\frac{1948}{105}\,\zeta_4\,\zeta_3^2+\frac{2608}{105}\,\zeta_2\,\zeta_5\,\zeta_3+\frac{12099}{224}\,\zeta_7\,\zeta_3\\
&\,+\frac{1335931}{47040}\,\zeta_5^2+\frac{25669936301}{63504000}\,\zeta_{10}\,, \\
f_{0,11} &\,= \frac{62}{315}\,\zeta_{5,3}\,\zeta_3+\frac{83}{120}\,\zeta_{5,3,3}-\frac{2872}{945}\,\zeta_2\zeta_3^3-\frac{13211}{672}\,\zeta_5\,\zeta_3^2-\frac{661411}{3024}\,\zeta_8\,\zeta_3\\
&\,-\frac{242776937}{725760}\zeta_{11}-\frac{605321}{3024}\,\zeta_5\,\zeta_6-\frac{2583643}{16200}\,\zeta_4\,\zeta_7-\frac{28702763}{340200}\,\zeta_2\,\zeta_9\,,\\
f_{0,12} &\,=\frac{74711}{162000}\zeta_{5,3}\,\zeta_4-\frac{13793}{7560}\,\zeta_{6,4,1,1}+\frac{3965011}{793800}\,\zeta_{7,3}\,\zeta_2-\frac{33356851}{4082400}\,\zeta_{9,3}\\
&\,+\frac{252163}{181440}\,\zeta_3^4+\frac{620477}{10080}\,\zeta_6\,\zeta_3^2+\frac{8101339}{75600}\,\zeta_4\,\zeta_5\,\zeta_3+\frac{342869}{3780}\,\zeta_2\,\zeta_7\,\zeta_3\\
&\,+\frac{101571047}{680400}\,\zeta_9\,\zeta_3+\frac{71425871}{1587600}\,\zeta_2\,\zeta_5^2+\frac{904497401571619}{620606448000}\,\zeta_{12}\\
&\,+\frac{484414571}{2721600}\,\zeta_5\,\zeta_7\,,\\
f_{0,13} &\,=
\frac{4513}{1890}\,\zeta_{5,3}\,\zeta_5+\frac{27248}{23625}\,\zeta_{5,3,3}\,\zeta_2-\frac{97003}{235200}\,\zeta_{5,5,3}+\frac{13411}{75600}\,\zeta_{7,3}\,\zeta_3\\
&\,+\frac{7997743}{12700800}\,\zeta_{7,3,3}-\frac{187318}{14175}\,\zeta_4\,\zeta_3^3-\frac{125056}{4725}\,\zeta_2\,\zeta_5\,\zeta_3^2-\frac{17411413}{302400}\,\zeta_7\,\zeta_3^2\\
&\,-\frac{5724191}{100800}\,\zeta_5^2\,\zeta_3-\frac{1874972477}{2376000}\,\zeta_{10}\,\zeta_3-\frac{2418071698069}{2235340800}\,\zeta_{13}\\
&\,-\frac{2379684877}{6048000}\,\zeta_{11}\,\zeta_2-\frac{297666465053}{523908000}\,\zeta_6\,\zeta_7-\frac{1770762319}{2494800}\,\zeta_5\,\zeta_8\\
&\,-\frac{229717224973}{628689600}\,\zeta_4\,\zeta_9\,.
\esp\eeq
The results have been reduced to a minimal set of multiple zeta values using the multiple zeta value data mine~\cite{Blumlein:2009cf}.
We have checked the values~(\ref{MNtotalcrossmore}) through 19 digits
by numerically evaluating eq.~\eqref{fnk} for $n=0$.

The Mueller-Navelet coefficients can again be obtained from a generating function similar to the one for the transverse momentum distribution,
\begin{equation}\label{eq:xsec_gen_func}
\hat{\sigma}_{gg} = \frac{\pi\,(C_A\as)^2}{2E_\bot^2}\,\Big\{Z(\eta) + 2\,\cP_{\zeta}\left[x_1\,\mathcal{X}(x_0,x_1;2\,\eta)\,\cZ(x_0;\eta)\right]\Big\}\,,
\eeq
where $\cP_\zeta$ is the linear map that sends a word to the corresponding multiple zeta value regularized by the shuffle multiplication~\cite{ShuffleReg} (cf. the definition of the Drinfel'd associator in Section~\ref{sec:SVHPLs}). We checked that the generating function reproduces the results of eq.~\eqref{MNtotalcrossmore}. The structure of the generating function can be understood as follows: The integral~\eqref{Ikdef} adds a ``1'' to every HPL in $B(\rho;\eta)$, which in terms of the non-commutative variables $\{x_0,x_1\}$ corresponds to multiplying by $x_1$ from the left. The regularized version of the integral~\eqref{Ikdef} is obtained by dropping all logarithmically divergent terms, i.e., by replacing all zeta values by their shuffle-regularized version. Finally, we need to partition the regularized version of the integral into its contribution from the Mueller-Navelet coefficients and the contribution from eq.~\eqref{Z_eta}.
The need to perform this partitioning can be seen by examining the first line
of eq.~\eqref{fixIde}, in which the terms at order $k$ with no powers of $\ln\delta$ are $f_{0,k} - Z_k$.  These are the terms generated by the shuffle
regularization.  Therefore we have to add the $Z(\eta)$ term in order to 
obtain eq.~\eqref{eq:xsec_gen_func}, the generating function for 
the Mueller-Navelet coefficients $f_{0,k}$.

Mueller and Navelet~\cite{Mueller:1986ey} also gave an asymptotic formula
for the behavior of $f_{0,k}$ as $k\rightarrow\infty$:
\beq\label{eq:MNapprox}
\sum_{k=0}^\infty \tilde{f}_{0,k} \eta^k =
\frac{1}{\pi\nu_0} \biggl[ \frac{1}{1+2\eta}
     - \frac{1}{8\nu_0^2 (1+\frac{2}{3}\eta)} \biggr] \,,
\eeq
where $\nu_0 = e^{-\gamma_E}$ and $\gamma_E$ is the Euler-Mascheroni constant.
In Table~\ref{tab:vsMNapprox}, we compare the numerical values
of the exact coefficients $f_{0,k}$ from eq.~\eqref{MNtotalcrossmore}
with the approximate values $\tilde{f}_{0,k}$ from this formula.
We see that by $k=13$ the exact and approximate values agree to 2 parts
in a billion.

\begin{table}[!h]
\begin{center}
\begin{tabular}{|r|c|c|}
\hline
\multicolumn{1}{|c|}{$k$} &\multicolumn{1}{c|}{$f_{0,k}$}
&\multicolumn{1}{c|}{$\tilde{f}_{0,k}$} \\
\hline
8 &  ~~145.13975384008912
  &  ~~145.12606589694502 \\
\hline
9 &  ~-290.25988683555143
  &  ~-290.26382715239066 \\
\hline
10 & ~~580.53650927568371
   & ~~580.53545121044840   \\
\hline
11 & -1161.07585293954502
   & -1161.07610035800818  \\
\hline
12 & ~2322.15572373880091
   & ~2322.15566600742394 \\
\hline
13 & -4644.31363149936796
   & -4644.31364220911962  \\
\hline
\end{tabular}
\caption{\label{tab:vsMNapprox} Numerical values of the exact coefficients
$f_{0,k}$ from eq.~\eqref{MNtotalcrossmore}, compared with the approximate
values $\tilde{f}_{0,k}$ for asymptotically large $k$ from
eq.~\eqref{eq:MNapprox}.}
\end{center}
\end{table}

\section{Conclusions}
\label{sec:concl}

In this paper, we have introduced the generating function
(\ref{eq:LLAeq}) which allows us to obtain, at each order in
perturbation theory, the coefficients of the leading logarithmic BFKL
Green's function in transverse momentum space.  In eq.~(\ref{fcoeffL})
we have explicitly shown the
coefficients of the first six loops of that $\as$ expansion.  This
allows us to exhibit analytically the dependence on the jet transverse
momenta of the dijet cross section in the large rapidity limit,
{\it i.e.}~the Mueller-Navelet jet cross section.  Accordingly, we have
provided fully analytic azimuthal-angle and transverse-momentum
distributions of the Mueller-Navelet jet cross section in terms of
harmonic polylogarithms.  We have also obtained the Mueller-Navelet 
total cross section through a generating function, and have computed 
its coefficients explicitly up to 13 loops.

It would be interesting to know whether the analysis presented above
can be extended to the BFKL Green's function at next-to-leading
logarithmic accuracy, either in QCD or in $\mathcal{N}=4$ super-Yang-Mills
theory, for
which we know that new classes of single-valued harmonic
polylogarithms will appear. That is left to future work.

\section*{Acknowledgments}

VDD is grateful to the Institut f\"ur Theoretische Physik,
Universit\"at Z\"urich, and to the CERN Theoretical Physics Unit for
the hospitality at the later stages of this work. This research was
supported by the Research Executive Agency (REA) of the European Union
through the Initial Training Network LHCPhenoNet under contract
PITN-GA-2010-264564, by the ERC grant ``IterQCD'', by the ERC grant 291377 ``LHCtheory: Theoretical predictions and analyses of LHC physics: advancing the precision frontier'', 
and by the US Department of Energy under contract DE-AC02-76SF00515.

\appendix


\section{A recursive formula for the Fourier coefficients $\chi_{\nu,n}$}
\label{sec:appa}

Using the recursive formula for the $\psi$ function
\beq
\psi(1+z) = \psi(z) + {1\over z}\, ,\label{recpsi}
\eeq
we obtain a recursive equation for $\chi_{\nu,n}$, 
\beq
\chi_{\nu,n+2} = \chi_{\nu,n} - {(1+n)\over
\nu^2 + \frac{(1+n)^2}4}\,,\qquad n \geq 0 \,. \label{recurs}
\eeq
By iterating it, we can make the $\chi$ function explicit for the 
even and odd modes,
\beqa
\chi_{\nu,2n} &=& \chi_{\nu,0} - \sum_{j=0}^{n-1}
{1+2j\over \nu^2 + \frac{(1+2j)^2}4}\,, \nonumber\\ 
\chi_{\nu,2n+1} &=& \chi_{\nu,1} - \sum_{j=0}^{n-1}
{2(j+1)\over \nu^2 + (j+1)^2}
\, ,\label{recurs2}
\eeqa
for $n \ge 1$. Note that we have in addition $\chi_{\nu,-n} = \chi_{\nu,n}$. Then one can substitute \eqn{recurs2} into \eqn{fnk}
and obtain a recursive formula for the Fourier coefficients.

\section{The small $\nu$ expansion of the BFKL eigenvalue $\chi_{\nu,n}$}
\label{sec:appb}

For small $\nu$, we can use the expansion 
\beq
\psi(1+z) -\psi(1)\, =\, \sum_{k=2}^{\infty} (-1)^k\, \zeta_k\, z^{k-1}\, ,
\eeq
valid for small $z$, and the doubling formula,
\beq
2\psi(2z) = 2\ln{2} + \psi(z)
 + \psi\left(z+\frac{1}{2}\right)\, ,\nonumber
\eeq
to obtain
\beq
\chi_{\nu,n}\, =\, 2\, \sum_{k=0}^{\infty} a_{kn} \nu^{2k}\,
,\label{omegacoeff}
\eeq
with coefficients
\beq
\begin{array}{ll}
a_{00} = 2\ln{2}\,, \quad & 
\quad a_{k0} = (-1)^k\, (2^{2k+1}-1)\, \zeta_{2k+1}\,,\\
a_{01} = 0\,, \quad & \quad a_{k1} = (-1)^k\, \zeta_{2k+1}\, .
\end{array} \quad \mbox{for}\quad k\ne 0\, .\label{zerouno}
\eeq
For $n=0$, that yields the usual expansion of the eigenvalue of the 
BFKL equation,
\beq
\chi_{\nu,0}\, =\, 2\, (2\ln{2} - 7\zeta_3 \nu^2 + \cdots )\,.
\label{chi0}
\eeq
Eq.~(\ref{recurs}) allows us to obtain a recursive formula for the
coefficients of \eqn{omegacoeff},
\beq 
a_{k,n+2} = a_{kn} + (-1)^{k+1} \left({2\over 1+|n|}\right)^{2k+1}\, .
\label{recurscoeff}
\eeq
For $k=0, 1$, \eqn{recurscoeff} yields
\beq
a_{0,n+2} = a_{0n} - {2\over 1+|n|} \qquad\quad a_{1,n+2} = a_{1n} +
\left({2\over 1+|n|}\right)^3\, ,\label{dueplus}
\eeq
in agreement with ref.~\cite{Stirling:1994he}. 

We can solve the formul\ae\ of \eqn{dueplus} explicitly, and write
\beq
\begin{array}{ll} \displaystyle
a_{0,2n} = a_{00} - \sum_{k=1}^n {2\over 2k-1} 
= \psi(1) - \psi\left(n+\frac{1}2\right)
\\ \displaystyle
a_{1,2n} = a_{10} + \sum_{k=1}^n \left({2\over 2k-1}\right)^3 
= \frac1{2}\ \psi''\left(n+\frac{1}2\right)
\\ \displaystyle
a_{0,2n+1} = a_{01} - \sum_{k=1}^n {1\over k} 
= \psi(1) - \psi\left(n+1\right)
\\ \displaystyle
a_{1,2n+1} = a_{11} + \sum_{k=1}^n \left({1\over k}\right)^3
= \frac1{2}\ \psi''\left(n+1\right)
\end{array} \qquad \mbox{with} \quad n\ge 1\, ,\label{fine}
\eeq
and with $a_{00}, a_{01}, a_{10}, a_{11}$ given in \eqn{zerouno}.

\bibliographystyle{JHEP}
\bibliography{refs}

\end{document}